\documentstyle[psfig]{mn}

\title[The Absorption Systems of GB1759+7539]{The Lyman-alpha Forest and Heavy Element Systems of GB1759+7539\footnotemark} 
\author[P.~J.~Outram et al.]{P.~J.~Outram$^1$, F.~H.~Chaffee$^2$,
R.~F.~Carswell$^1$\\ $^1$ Institute of Astronomy,
Madingley Road, Cambridge CB3 0HA\\ $^2$ W. M. Keck Observatory,
65-1120 Mamalahoa Hwy., Kamuela, HI 96743, U.S.A.\\ }

\begin{document}
\maketitle
\begin{abstract}
We present observations of the high-redshift QSO GB1759+7539
($z_{\rm em}=3.05$) obtained with HIRES on the Keck 10m telescope. The
spectrum has a resolution of FWHM = 7 km$\,$s$^{-1}$, and a typical signal-to-noise ratio per 2 km$\,$s$^{-1}$ pixel of
$\sim$ 25 in the Ly$\alpha$\  forest region, and $\sim$ 60 longward
of the Ly$\alpha$\  emission.

The observed Ly$\alpha$\  forest systems have a mean redshift of
$<z>$=2.7.  The H$\:${\small I} column density distribution is well
described by a power law distribution with index $\beta = 1.68 \pm
0.15$ in the range $13.5 <$ log$ N < 14.5$. The Doppler width
distribution is consistent with a Gaussian distribution of mean
$b=26$ km$\,$s$^{-1}$, and standard deviation $\sigma=12$ km$\,$s$^{-1}$ with a
cut-off at $b=20$ km$\,$s$^{-1}$. There is marginal evidence of clustering
along the line of sight over the velocity range 100$<\Delta
v<$250 km$\,$s$^{-1}$. The 1-point and 2-point joint probability
distributions of the transmitted flux for the Ly$\alpha$\  forest were
calculated, and shown to be very insensitive to the heavy element
contamination. We could find no evidence of Voigt profile departures
due to infalling gas, as observed in the simulated forest spectra.

Twelve heavy-element absorption systems were identified, including
damped Lyman-alpha (DLA) systems at $z_{\rm abs}$=2.62  and 2.91. The  C,
N, O, Al, Si, P, S, Mg, Fe, and Ni absorption features of these
systems were studied, and the elemental abundances calculated for the
weak unsaturated lines. The systems have metallicities of
$Z_{\rm 2.62}\simeq1/20 Z_{\rm \odot}$ and $Z_{\rm 2.91}\simeq1/45
Z_{\rm \odot}$. Both systems appear to have a low dust content. They show
an over-abundance of $\alpha$-elements relative to Fe-peak elements,
and an under-abundance of odd atomic number elements relative to
even. Nitrogen was observed in both systems, and found to be
under-abundant relative to oxygen, in line with the time delay model of
primary nitrogen production. C$\:${\small II}* was also seen, allowing us to determine an
upper limit to the cosmic microwave background temperature at $z=2.62$
of T$_{\rm CMB} <$12.9K.

\end{abstract}

\section{Introduction}
\footnotetext{Based on observations obtained at the W. M. Keck Observatory, which is jointly operated by the University of California and the California Institute of Technology.}
In recent years rapid progress has been made in observational and
theoretical studies of Ly$\alpha$ absorption lines in the spectra of
high-redshift QSOs. Large telescopes have given us the opportunity of
studying the lines in much greater detail than ever before, with large
gains in both resolution and signal-to noise ratio (S/N) (e.g. Hu et al. 1995; Kirkman and Tytler 1997). At the same
time, recent cosmological simulations, taking into account a
photoionizing ultra-violet background, have suggested that the
Ly$\alpha$ clouds at high redshift develop naturally in a hierarchical
structure-formation scenario (e.g. Cen et al. 1994; Miralda-Escud\'e et al. 1996). The detailed
study of the Ly$\alpha$ forest, and comparison with the results of
such simulations is perhaps one of the best tests of cosmic
structure-formation theories, providing a constraint in the era
$2<z<5$.

The heavy element systems observed in the line of sight to QSOs also provide a
wealth of information about the formation of structure at high
redshift. Understanding the chemical evolutionary history of galaxies
is fundamental to the study of galaxy formation. The largest heavy element
systems, damped Lyman-alpha systems (DLAs), are believed to be the
progenitors of present day galaxies \cite{wol95}. They dominate the
mass of neutral gas at redshift $z\simeq3$, a mass comparable with
that of the stars in present day spiral disks, suggesting that DLAs
are the source of most of material available for star formation at
high redshift \cite{lan95}

Prochaska \& Wolfe (1997) studied the kinematics of DLAs, concluding
that there is evidence for rotation, supporting the theory that they
are the progenitors of modern day disk galaxies. This result was
disputed by Haehnelt, Steinmetz \& Rauch (1997). CDM models infer that
DLAs are more like the progenitors of dwarf galaxies, or galactic
halos, a model that is backed by studies of heavy element abundances
\cite{max94,lu96a} which have found an abundance pattern more akin
to halo stars than disk stars in our Galaxy.

Pettini et al.\ (1994) systematically studied the Zn, and Cr
abundances in DLAs, and concluded that they had low metallicity,
typically $Z_{\rm DLA}\simeq1/10 Z_{\rm \odot}$ at $z\simeq2$, and that they
have a much lower dust-to-gas ratio than the local interstellar medium.
 Although it is
still unclear which population of galaxies gives rise to DLAs, it is
apparent that the gas in DLAs is at an early stage in the chemical
evolution of the object, and so abundance studies should give us an
insight into the early stages of galaxy evolution.

In this paper, we present observations and analyses of the absorption
spectrum of the QSO GB1759+7539 ($z_{\rm em}=3.05$). In \S 2 we describe
the observations and data reduction methods. The analysis of the
absorption lines, and identification of the heavy element lines is discussed
in \S 3. Then, in \S 4, we present the main observational results
concerning the Ly$\alpha$\  forest, including the column density, and
Doppler width distributions, and the clustering properties, as
indicated by the two point correlation function. The elemental
abundances in the two damped systems are discussed in \S 5, where the
dust content, and nucleosynthesis histories of the two objects are
considered. \S 6 investigates the cosmic microwave background
temperature at $z=2.62$. The main conclusions are summarized in \S 7.

\section{Observations and Data Reduction}
The radio source, GB1759+7539 was selected from the Green Bank 5-GHz
survey \cite{con89}, and identified by Hook et al.\ (1996) as
a high-redshift $z_{\rm em}=3.05$ radio-loud QSO. Hook et al.\ 
noted that this object is optically very bright and hence it is ideal
for high resolution spectroscopic study.

\begin{table}
\caption{\rm Journal of Observations}
\label{tbl-1}
\begin{tabular}{ccc}
\hline   Dates  & Exposure time (s) &  Wavelength range(\AA)\\
 \hline   6 July 1997  & 9000  & 4115--6515  \\ 
6 July 1997  & 6000  & 4140--6540
 \\ 13 July 1997  & 6000  & 4115--6515\\ 
13 July 1997  & 9000  & 4140--6540\\
 \hline
\end{tabular}
\end{table}

The data used in the present study were obtained in two nights (see
Table 1) using the High Resolution Echelle Spectrometer (HIRES) on the
10m Keck telescope \cite{vog92}, with the TeK 2048x2048 CCD. The
FWHM of the instrument profile was found to be about 7 km$\,$s$^{-1}$. The
HIRES setup is such that complete spectral coverage is only possible
for $ \lambda < 5100$\AA, so we used two partially overlapping setups
to obtain complete coverage over the wavelength range 4116 - 6540
\AA. This leads to fairly large variations in the S/N from region to
region in the final spectrum.

Each image was bias and flat-field corrected using IRAF routines. The
cosmic rays were flagged using a median filter and given zero weight
in the individual frames. The sky-subtracted optical spectra were then
optimally extracted, along with a one-sigma error estimate, calibrated
to vacuum heliocentric wavelengths, and flux calibrated.

Even after flux calibration, there were some inexplicable low-order
variations in the flux level of some of the images. The variations
were echelle order dependent, and therefore clearly instrument
related. These were removed by picking an apparently unaffected frame
as a template, fitting low-order polynomials to the ratio of each
frame to the template, and dividing out the variations. This procedure
also scaled each spectrum to the same flux level. As the resulting
spectrum was used solely for absorption line studies, where the ratio
of the line intensity to that of the continuum is important, but the
actual flux is not, this should have little effect on any of the
analysis.

The echelle orders were then resampled to the same dispersion, and
added together weighted according to their S/N. Finally, any bad
pixels that escaped attention earlier were corrected and flagged to
get zero weight in the line fitting routines. Atmospheric molecular
oxygen absorption was removed from the region 6280 - 6310\AA\  by
dividing by a template generated from a standard star observed at
similar airmass.

The continuum level redward of the Ly$\alpha$\  emission line was
estimated by fitting cubic splines to regions free from absorption
lines, using the IRAF continuum fitting routine. The continuum for the
Ly$\alpha$\  forest region was fitted using the small regions deemed
to be free of absorption, interpolating between these regions with a
low-order polynomial fit. The region containing O$\:${\small VI} and Ly$\beta$\
emission was fitted separately, in a similar manner, but with a higher
order polynomial fit. The resulting continuum appears to fit the data
well; however, in regions where the fitted continuum was possibly not
accurate, it was allowed to vary in the line fitting routine VPFIT
(see later.) These variations were never found to be more than
about 2\% of the original fitted continuum level.
\section{Data Analysis}

Voigt Profiles were fitted to the absorption lines, using the software 
package VPFIT \cite{web87,coo94}, in order to determine the redshifts, column
densities and Doppler widths of ions with observed absorption lines.

The procedure uses a reduced $\chi^2$ technique, which adjusts the
parameters of an initial guess in order to minimize the $\chi^2$
value. The spectrum was fitted in sections, using the smallest regions
possible, bounded by where the spectrum reaches the continuum
level. After an initial guess, further lines were automatically added
until the addition of extra components failed to significantly
reduce the normalised $\chi^2$ further (as described in Rauch et
al. 1992). This usually resulted in a normalised $\chi^2 \sim
1.1$. Occasionally, such a good fit was not quite possible, due to
narrow non-Gaussian noise spikes in the spectrum; probably caused by 
CCD defects, or cosmic rays not fully removed in the data reduction process.

In some spectral regions where the reduced $\chi^2$ was greater than $\sim
1.1$ the VPFIT program attempted to reduce it further by adding weak narrow features fitted to what are evidently noise spikes. A feature of the fits to these features is that the parameter error estimates are large, so they were easily identifiable. To avoid the possibility of overfitting in this way, such features were removed and the spectral region refitted.

The Voigt profile fitting procedure does not necessarily give unique
results (as noted by Kirkman \& Tytler 1997.) Often from a slight
change in the initial guess, the routine settles on different sets of
Voigt profiles which both fit the same absorption feature, and
satisfy the criteria for a satisfactory fit. The
intrinsic errors quoted in this paper are just the formal parameter
fitting errors, assuming that the fitted solution in terms of Voigt
profiles is correct. Different fits can often yield different
solutions with differences much greater than this intrinsic
error. This is particularly noticeable for saturated lines, where
formal errors in column density can be less than 0.1 dex, but an extra
line can alter the fitted column density by 2 dex or more due to the
position of the feature on the curve of growth. Therefore, in general,
the column density of saturated lines is ignored in this paper, or a
very generous lower limit is used.

All fitted lines in the forest were initially assumed to be
Ly$\alpha$. The heavy element systems longward of the Ly$\alpha$\
emission line were analysed and all but three absorption lines were
positively identified. The Ly$\alpha$\  forest was then searched for
any new lines belonging to known heavy element systems, before it was finally
searched for new heavy element systems. All lines that were not positively
identified as a heavy element absorption line have been identified as
Ly$\alpha$\  for the purposes of this paper.

Atomic data, including oscillator strength, rest-frame vacuum wavelength and radiation damping constant are taken from Morton (1991), with revised values from Tripp, Lu, \& Savage (1996). We have adopted recent oscillator strength determinations for Ni$\:${\small II} (Fedchak \& Lawler 1999; Zsarg\'o \& Federman 1998), and the weak Mg$\:${\small II} transitions (Fitzpatrick 1997). Ni$\:${\small II}$\;\lambda 1317$ remains uncertain, and so this line was not used as a constraint.

\subsection{Heavy element absorption systems}
A total of twelve heavy element absorption systems were identified,
ranging from a single C$\:${\small IV} doublet, to the complex absorption systems
at $z_{\rm abs}=2.625$ and $z_{\rm abs}=2.911$. Apart from the interest in
the heavy element systems themselves, the identification of heavy element lines is
crucial to minimise the contamination of the sample of Ly$\alpha$\
lines with misidentified heavy element lines in order to study  the
Ly$\alpha$\  forest. Below is a brief summary of the heavy element
absorption systems observed.\\

\centerline{\bf{$z_{\rm abs}=0.000$ (3 components)}}

Galactic absorption of Na$\:${\small I} $\lambda\lambda 5891,5897$ was
observed. Three components were fitted, with blueshifts of 16 - 46
 km$\,$s$^{-1}$ relative to the heliocentric rest-frame. A 14 km$\,$s$^{-1}$ shift redwards must be applied to correct for the motion of the Sun, and obtain the local standard of rest velocity. One component is almost stationary in this frame, and therefore probably is absorption from the local interstellar medium.\\

\centerline{\bf{$z_{\rm abs}=1.348$ (3 components)}}

Three components of Fe$\:${\small II}, $\lambda2344, 2374, 2382, 2586,$ and $2600$, at redshifts $z_{\rm abs}=1.3479, 1.3481,$ \& $1.3485$, were observed
longward of the Ly$\alpha$\  emission
line. Al$\:${\small III}$\lambda\lambda1854, 1862$, the only other potentially observable heavy element line within the range covered, lies in the
Ly$\alpha$\  forest and is obscured by other absorption lines.\\

\centerline{\bf{$z_{\rm abs}=1.8848$}}

A single narrow-component C$\:${\small IV} doublet $\lambda\lambda 1548,1550$ 
was clearly detected at $z_{\rm abs}=1.8848$ in the Ly$\alpha$\
forest. No other heavy element lines were detected in this system.\\

\centerline{\bf{$z_{\rm abs}=1.935$ (complex)}}

There is a complex heavy element system at $z_{\rm abs}=1.935$, with C$\:${\small IV}, Si$\:${\small II},
Si$\:${\small IV}, Al$\:${\small II}, and Al$\:${\small III} absorption detected. The C$\:${\small IV}$\lambda\lambda
1548,1550$ absorption feature requires six components and has
components at velocities of $-150$ and +180 km$\,$s$^{-1}$ relative to the
central ones. It lies in the Ly$\alpha$\  forest, and is blended with
S$\:${\small II} $\lambda1253$ at $z_{\rm abs}=2.625$ and Si$\:${\small II} $\lambda1304$ at $z_{\rm abs}=2.484$ as well as Ly$\alpha$\  forest lines, so some
confusion is possible. Si$\:${\small IV} $\lambda1402$  , but not
$\lambda1393$   lies within the wavelength range observed, and has
a similar structure to that of the C$\:${\small IV}
absorption. Si$\:${\small II} $\lambda1526$  , and Al$\:${\small II} $\lambda1670$   both
lie in the Ly$\alpha$\  forest as well, but
Al$\:${\small III} $\lambda\lambda1854, 1862$   was observed redward of the
Ly$\alpha$\  emission line. Absorption from these three
lower-ionization lines was detected only from the central component.\\
   
\centerline{\bf{$z_{\rm abs}=2.4390$}}

A single C$\:${\small IV} doublet $\lambda\lambda 1548,1550$   was clearly
detected at $z_{\rm abs}=2.4390$, longward of the Ly$\alpha$\  emission
line. This corresponds to a Ly$\alpha$\  line which, if it is also
single, has column density logN(H$\:${\small I})$=14.6$. No
other heavy element lines were detected for this system.\\

\centerline{\bf{$z_{\rm abs}=2.484$ (complex)}}

The $z_{\rm abs}=2.484$ system has a complicated absorption structure. The
Ly$\alpha$\  feature has four main components, two at around
$z_{\rm abs}=2.484$, and outlying components at around $\pm250$ km$\,$s$^{-1}$,
all of which are saturated. The central components have strong
low-ionization absorption lines:
Si$\:${\small II} $\lambda1190, 1193, 1260, 1304, 1526$, O$\:${\small I} $\lambda1302$,
Al$\:${\small II} $\lambda1670$, C$\:${\small II} $\lambda1334$  , and
Fe$\:${\small II} $\lambda1608$   as well as the higher ionization ions Si$\:${\small III},
Si$\:${\small IV}, C$\:${\small IV}, whereas the outer systems only show absorption in the
latter ions. C$\:${\small IV} $\lambda1548$ is heavily blended with
Si$\:${\small IV} $\lambda1402$ at $z_{\rm abs}=2.84$. Al$\:${\small II} $\lambda1670$,
Si$\:${\small II} $\lambda1526$, and Fe$\:${\small II} $\lambda1608$   all lie in clear
parts of the spectrum and so are well constrained; however, the other
lines all lie in the forest where confusion and blending are more
likely. C$\:${\small II} $\lambda1334$   and Si$\:${\small IV} $\lambda\lambda 1393, 1402$ 
in particular are heavily blended with Ly$\alpha$\ .\\

\centerline{\bf{$z_{\rm abs}=2.625$ (complex)}}

The damped Ly$\alpha$\  system has a column density of
logN(H$\:${\small I})$=20.761\pm0.007$. Redward of the Ly$\alpha$\
emission line, Si$\:${\small IV} $\lambda\lambda 1393, 1402$,
Si$\:${\small II} $\lambda1526$, C$\:${\small IV} $\lambda\lambda 1548, 1550$,
Fe$\:${\small II} $\lambda\lambda 1608, 1611$  , Al$\:${\small II} $\lambda1670$  and Ni$\:${\small II} $\lambda=1751, 1741, 1709, 1454,$ and $1370$   were observed. The
Al$\:${\small II} $\lambda1670$   feature was blended with C$\:${\small IV} $\lambda1548$ at
$z_{\rm abs}=2.91$. The high ionization lines, C$\:${\small IV} $\lambda\lambda
1548, 1550$, and Si$\:${\small IV} $\lambda\lambda 1393, 1402$   show a single
sharp component at $z_{\rm abs}=2.6216$, also seen in C$\:${\small II} $\lambda1334$ , 200 km$\,$s$^{-1}$ blueward of the main system, which is fitted by seven Voigt profiles.

Upon searching the Ly$\alpha$\  forest, further absorption lines; N$\:${\small I} $\lambda=1199.5, 1200.2,$ and $1200.7$  , N$\:${\small V} $\lambda\lambda
1238, 1242$  , Mg$\:${\small II} $\lambda\lambda 1239, 1240$, S$\:${\small II} $\lambda=1250, 1253,$ and $1259$, P$\:${\small II} $\lambda1152$,
Si$\:${\small III} $\lambda1206$, O$\:${\small I} $\lambda1302$, and
C$\:${\small II} $\lambda1334$   were detected.  Si$\:${\small II} $\lambda=1190, 1193, 1260,$ and $1304$  were detected in the
Ly$\alpha$\  forest and fitted simultaneously with the other lines
from the same ion species redward of the Ly$\alpha$\  emission. Ni$\:${\small II} $\lambda1317$ was also seen, but not used simultaneously in the fit with the other Ni$\:${\small II} lines due to uncertainty in its oscillator strength. The
N$\:${\small I} $\lambda=1199.5, 1200.2,$ and $1200.7$   absorption features
were very distinct and virtually free from Ly$\alpha$\
blending. Mg$\:${\small II} $\lambda\lambda 1239, 1240$  , on the other hand, was blended with H$\:${\small I}, and so the column density derived can
only be taken as an upper limit.

\begin{table}
\caption{Measurements for the $z_{\rm abs}$=2.62 System}
\label{tbl-2}

\begin{tabular}{cccc}
\hline   Ion &log N$\pm\sigma$ & notes &[Z/H]\\ \hline   H I  &
20.761$\pm$0.007 & & \\ C$\:${\small II}*
& 12.808$\pm$0.063 &in forest & $<$-0.81$^a$  \\ C$\:${\small IV}  &
14.626$\pm$0.005 & &  \\ N$\:${\small I}  & 14.985$\pm$0.025 &in forest &
-1.83$\pm$0.03\\   N$\:${\small V}  & 13.429$\pm$0.035  &heavily blended  & \\  Mg$\:${\small II}  & 15.721$\pm$0.059 &heavily
blended  & $<$-0.57 \\ Si$\:${\small IV}  & 14.189$\pm$0.005 & &   \\ P$\:${\small II}  &
13.179$\pm$0.068$^b$ &in forest  & -1.16$\pm$0.07 \\ S$\:${\small II}  &
15.212$\pm$0.014 &in forest  & -0.82$\pm$0.02 \\ Fe$\:${\small II}  &
14.936$\pm$0.009 &  & -1.34$\pm$0.01 \\ Ni$\:${\small II}  & 13.889$\pm$0.007 &  &
-1.12$\pm$0.01  \\ \hline
\end{tabular}

\medskip
$^a$ Assuming T$_{\rm CMB}\propto(1+z)$\\ $^b$ Adopted value; corrected
for Ly$\alpha$\  obscuration.\\
\end{table}

\begin{figure}
\centerline{\hbox{\psfig{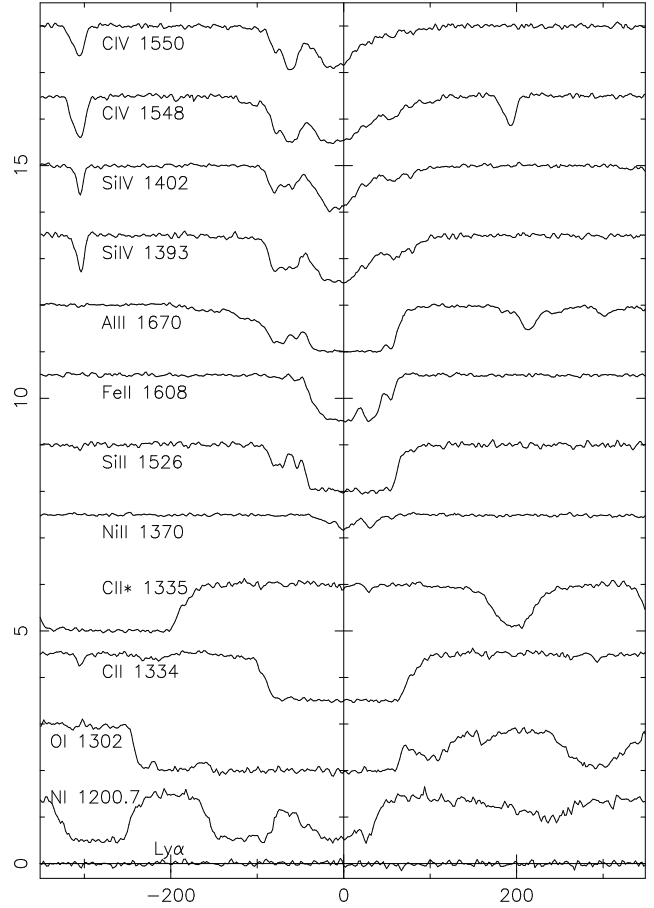}}}
\caption{Velocity profiles (km$\,$s$^{-1}$) of heavy element lines in the $z_{\rm abs}=2.625$ DLA. The spectra are all normalized to unit continuum. The zero velocity is fixed at $z=2.62528$. The O$\:${\small I} line lies in the forest, and is blended with Si$\:${\small III} from the $z=2.91$ system. Blueward of the N$\:${\small I}$\;\lambda 1200.7$ line, at $v=-250$ and $-125$ km$\,$s$^{-1}$ can be seen the other two lines in the triplet; N$\:${\small I}$\;\lambda\lambda 1199.5, 1200.2$. }
\label{figv1}
\end{figure}

 The main component of P$\:${\small II}($z_{\rm abs}=2.6252$) was clearly detected, but
 any components at a slightly higher redshift are blanketed by a
 Ly$\alpha$\  line. Assuming, however, that the P$\:${\small II} feature has the
 same shape as the other low ionization lines, the other components
 would have a column density of half a dex less than the main
 component observed. Correcting for this obscuration leads to an
 increase in the abundance of Phosphorus by 0.24 dex. S$\:${\small II} $\lambda1253,$ and $1259$   are blended, but S$\:${\small II} $\lambda1250$ is
 not, allowing an accurate determination of the abundance of
 Sulphur. Many of the heavy element lines were saturated, but amongst the low
 ionization species, abundances of S$\:${\small II}, Fe$\:${\small II}, Ni$\:${\small II}, N$\:${\small I}, and P$\:${\small II} were
 determined. Excited C$\:${\small II}* $\lambda1335$ was also marginally detected. The velocity profiles of many of the absorption features detected for this system can be seen in figure~\ref{figv1}.\\

\centerline{\bf{$z_{\rm abs}=2.7871$}}

A single C$\:${\small IV} doublet $\lambda\lambda 1548, 1550$ was clearly
detected at $z_{\rm abs}=2.7871$, longward of the Ly$\alpha$\  emission
line. This corresponds to a Ly$\alpha$\  line with column density
logN(H$\:${\small I})$=14.5$.\\

\centerline{\bf{$z_{\rm abs}=2.795$ (2 components)}}

Two C$\:${\small IV} doublets $\lambda\lambda 1548, 1550$   were detected at
$z_{\rm abs}=2.795,$ \& $2.796$, corresponding to a Ly$\alpha$\  line with
column density logN(H$\:${\small I})$=16.2$.\\

\centerline{\bf{$z_{\rm abs}=2.835$ (complex)}}

C$\:${\small IV} $\lambda\lambda 1548, 1550$  (five components) and
Si$\:${\small IV} $\lambda\lambda 1393, 1402$   (one component) were observed at
$z_{\rm abs}=2.835$. The corresponding Ly$\alpha$\  line has a column
density of logN(H$\:${\small I})$=15.7$.\\

\centerline{\bf{$z_{\rm abs}=2.84$ (complex)}}

C$\:${\small IV} $\lambda\lambda 1548, 1550$  absorption spanning a total
velocity interval of 350 km$\,$s$^{-1}$, with 16 components, was observed
at $z_{\rm abs}=2.84$. C$\:${\small II} $\lambda1334$, Si$\:${\small III} $\lambda1206$, and
Si$\:${\small IV} $\lambda\lambda 1393, 1402$   were also detected, showing a
similar distribution. The saturated Ly$\alpha$\  line has a measured
column density of logN(H$\:${\small I})$=18.3$.\\
  
\centerline{\bf{$z_{\rm abs}=2.896$ (3 components)}}

Three C$\:${\small IV} $\lambda\lambda 1548, 1550$  doublets were observed at
$z_{\rm abs}=2.8954, 2.8956,$ \& $2.8973$, corresponding to two
Ly$\alpha$\  clouds of column density logN(H$\:${\small
I})$=14.0$ \& $14.2$.\\

\centerline{\bf{$z_{\rm abs}=2.910$ (complex)}}

\begin{figure}
\centerline{\hbox{\psfig{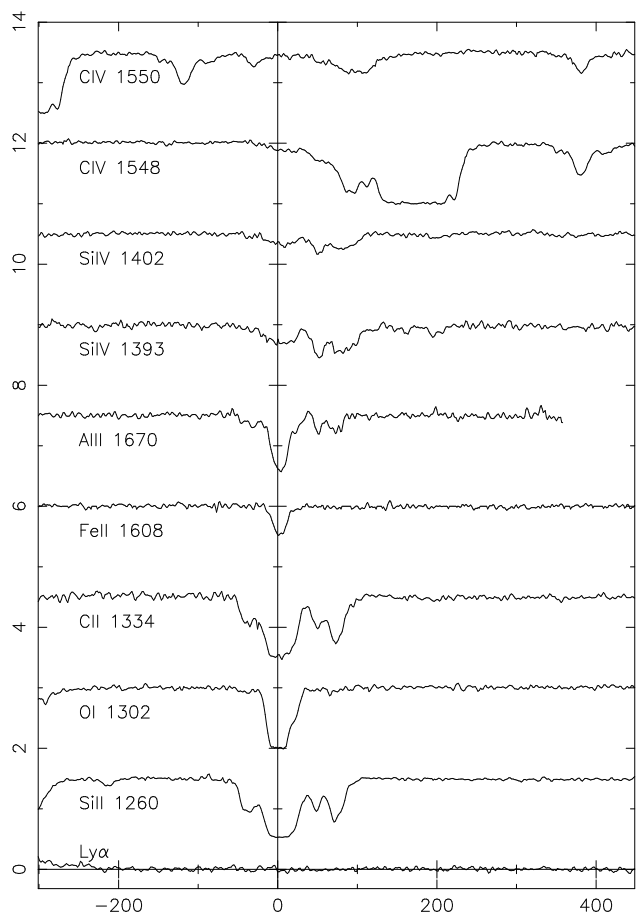}}}
\caption{Velocity profiles (km$\,$s$^{-1}$) of heavy element lines in the $z_{\rm abs}=2.91$ system. The spectra are all normalized to unit continuum. The zero velocity is fixed at $z=2.91016$. The saturated absorption feature partially obscuring C$\:${\small IV}$\;\lambda 1548$ is Al$\:${\small II} from the DLA system. The Al$\:${\small II} feature for this system lies near the edge of the observed spectrum.}
\label{figv2}
\end{figure}

The hydrogen column density of this large system is
logN(H$\:${\small I})$=19.795\pm0.006$. There is a second Ly$\alpha$\
component with logN(H$\:${\small I})$=17.021\pm0.205$ lying
420 km$\,$s$^{-1}$ distant from the main system. Only
C$\:${\small IV} $\lambda\lambda 1548, 1550$ was detected for this
component. C$\:${\small II} $\lambda1334$, C$\:${\small IV} $\lambda\lambda 1548, 1550$, O$\:${\small I} $\lambda1302$, Al$\:${\small II} $\lambda1670$, Si$\:${\small II} $\lambda=1304,$ and $1526$, Si$\:${\small IV} $\lambda\lambda 1393, 1402$, and Fe$\:${\small II} $\lambda1608$   absorption features were all detected longwards
of the Ly$\alpha$\  emission line, and abundances were calculated for
the unsaturated lines. Si$\:${\small II} $\lambda=1190, 1193,$ and $1260$  were
detected in the forest and fitted, by seven components, simultaneously
with the Si$\:${\small II} lines redward of the Ly$\alpha$\  emission. Si$\:${\small III} $\lambda1206$, N$\:${\small I} $\lambda=1199.5, 1200.2,$ and $1200.7$, and N$\:${\small V} $\lambda\lambda 1238, 1242$  were tentatively seen in the Ly$\alpha$\  forest, however the lines were very heavily
blended in each case. S$\:${\small II} $\lambda 1259$ was also seen, but S$\:${\small II} $\lambda=1250,$ and$ 1253$ were obscured in the forest. The velocity profiles of many of the absorption features identified from this system can be seen in figure~\ref{figv2}.
 
\begin{table}
\caption{Measurements for the $z_{\rm abs}$=2.91 System}
\label{tbl-3}

\begin{tabular}{cccc}
\hline   Ion &log N$\pm\sigma$ & notes &[Z/H]\\ \hline   H I  &
19.795$\pm$0.006 & &  \\ C$\:${\small II}  & 14.539$\pm$0.023 & & -1.82$\pm$0.02\\
C$\:${\small IV}  & 13.852$\pm$0.010 & &   \\ 
N$\:${\small I}  &  13.242$\pm$0.035  &heavily blended  & $<$-2.45 \\
N$\:${\small V}  & 12.758$\pm$0.116  &heavily
blended  &  \\ Al$\:${\small II}
& 12.764$\pm$0.008 & & -1.51$\pm$0.01  \\ Si$\:${\small II}  & 14.083$\pm$0.007 &
& -1.26$\pm$0.01 \\ Si$\:${\small III}  & 14.541$\pm$0.262 &heavily blended  &  \\
Si$\:${\small IV}  & 13.479$\pm$0.005 & &    \\
S$\:${\small II}  & 13.674$\pm$0.023 & & -1.39$\pm$0.02  \\ Fe$\:${\small II}  & 13.658$\pm$0.010 &  &
-1.65$\pm$0.01\\ \hline
\end{tabular}
\end{table}

\vspace{2mm}

\begin{figure*}
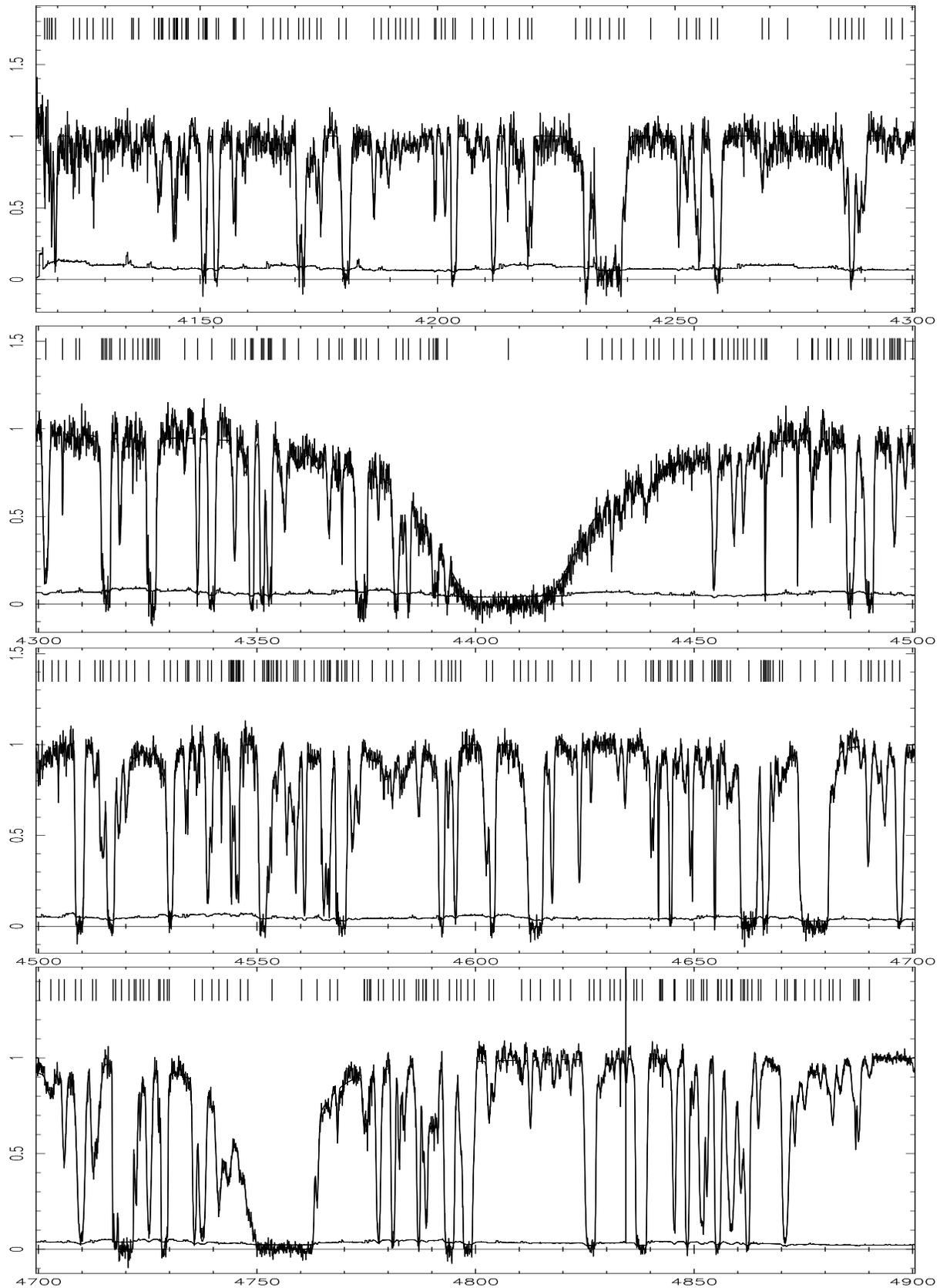

\begin{minipage}{170mm}
\centerline{\hbox{\psfig{figure=plot1.ps,height=5.5cm,width=16cm,angle=270}}}
\centerline{\hbox{\psfig{figure=plot2.ps,height=5.5cm,width=16cm,angle=270}}}
\centerline{\hbox{\psfig{figure=plot3.ps,height=5.5cm,width=16cm,angle=270}}}
\centerline{\hbox{\psfig{figure=plot4.ps,height=5.5cm,width=16cm,angle=270}}}
\caption{The spectrum of GB1759+7539, plotted against vacuum
heliocentric wavelength(\AA), normalized to unit continuum, together
with overlying profile fits. The tick marks show the position of the
components given in Table 4. The 1$\sigma$ error is also shown.}
\label{fig1}
\end{minipage}
\end{figure*}

A total of 725 lines were fitted, of which 322 are Ly$\alpha$\  forest
lines, 249 are heavy element lines redward of the Ly$\alpha$\
emission (246 positively identified) and 152 are identified heavy
element lines within the Ly$\alpha$\  forest. The continuum-normalized spectrum,
plotted against vacuum heliocentric wavelength(\AA), together with
overlying profile fits, and the 1$\sigma$ error are shown in
figure~\ref{fig1}. There are a number of spikes in the error due to
cosmic rays or CCD defects. The error spikes around 6280\AA\  are due
to molecular oxygen absorption that was removed from the spectrum. The
tick marks indicate the positions of line features. A list of the line
identifications, together with the fitted parameters is given in
table~\ref{tbl-4}.

\setcounter{figure}{2}
\begin{figure*}
\begin{minipage}{170mm}
\centerline{\hbox{\psfig{figure=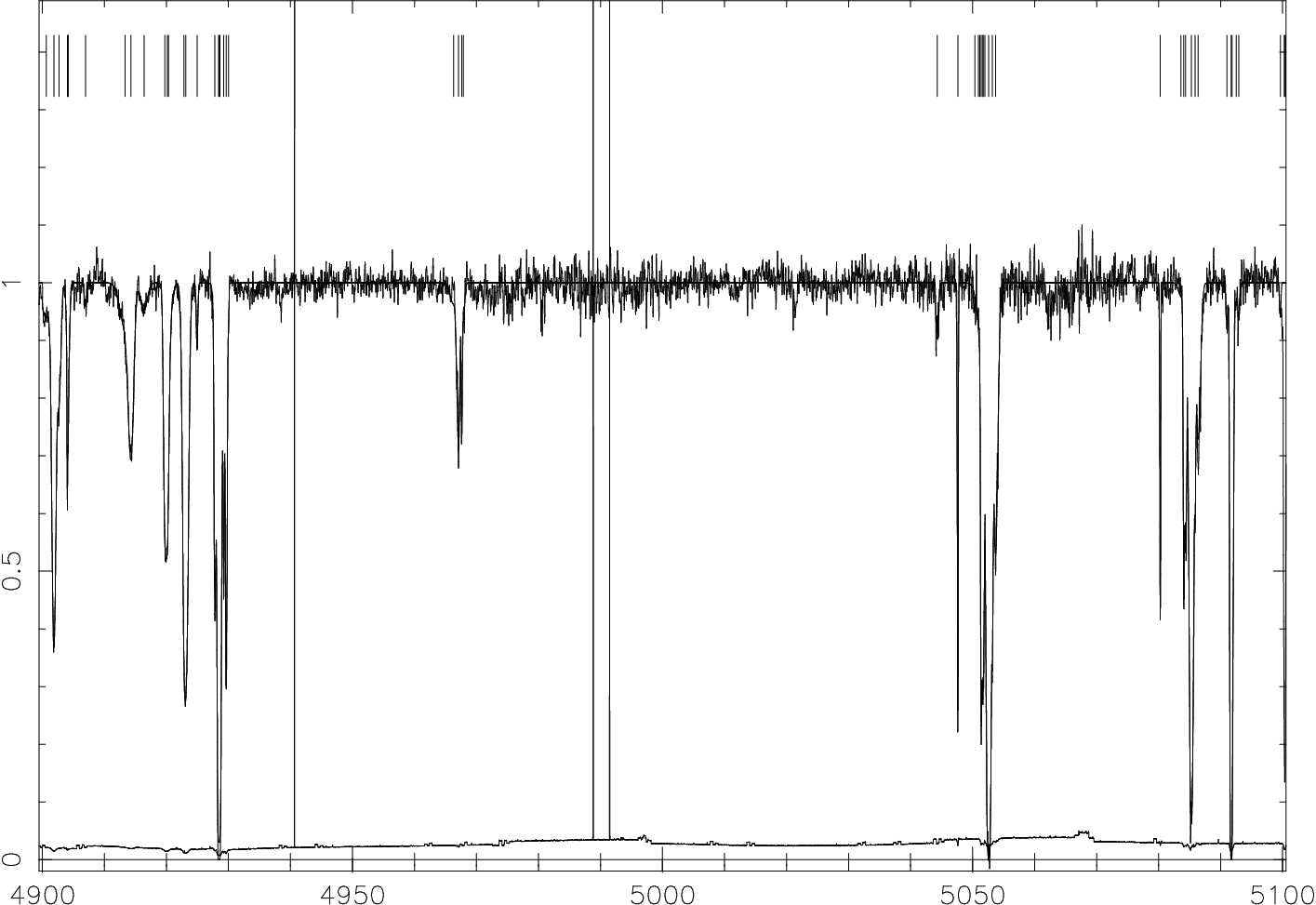,height=5.5cm,width=16cm,angle=270}}}
\centerline{\hbox{\psfig{figure=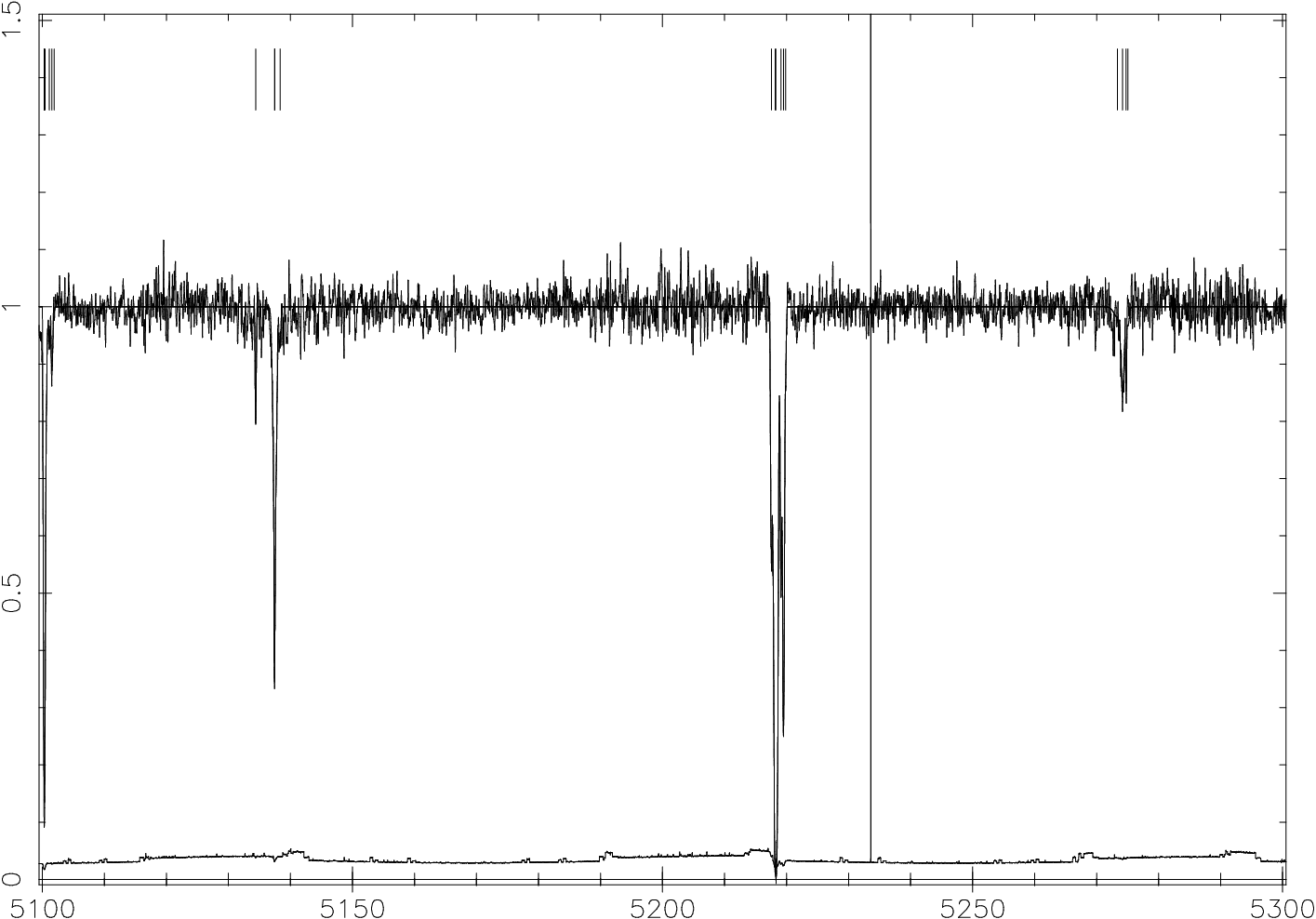,height=5.5cm,width=16cm,angle=270}}}
\centerline{\hbox{\psfig{figure=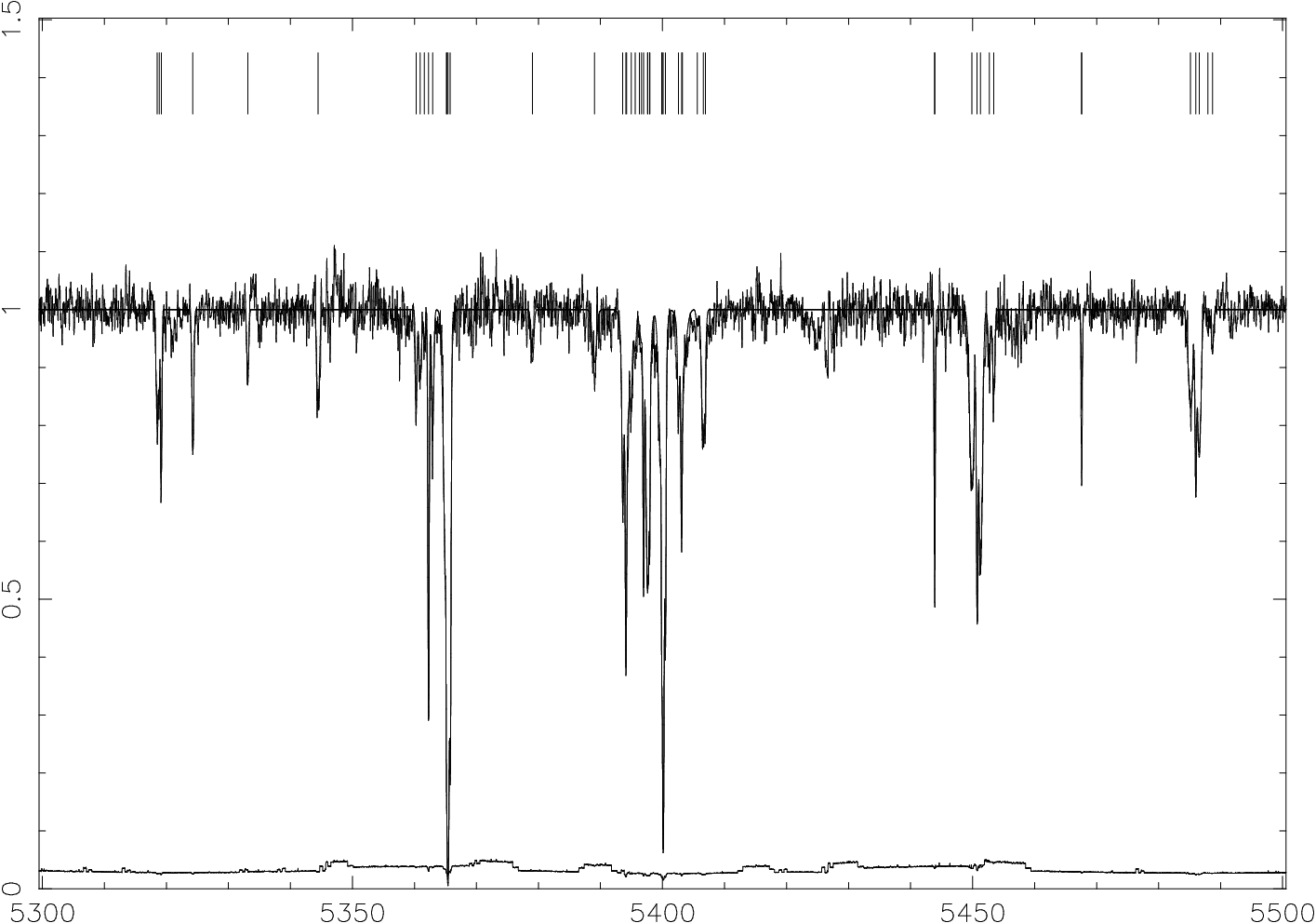,height=5.5cm,width=16cm,angle=270}}}
\centerline{\hbox{\psfig{figure=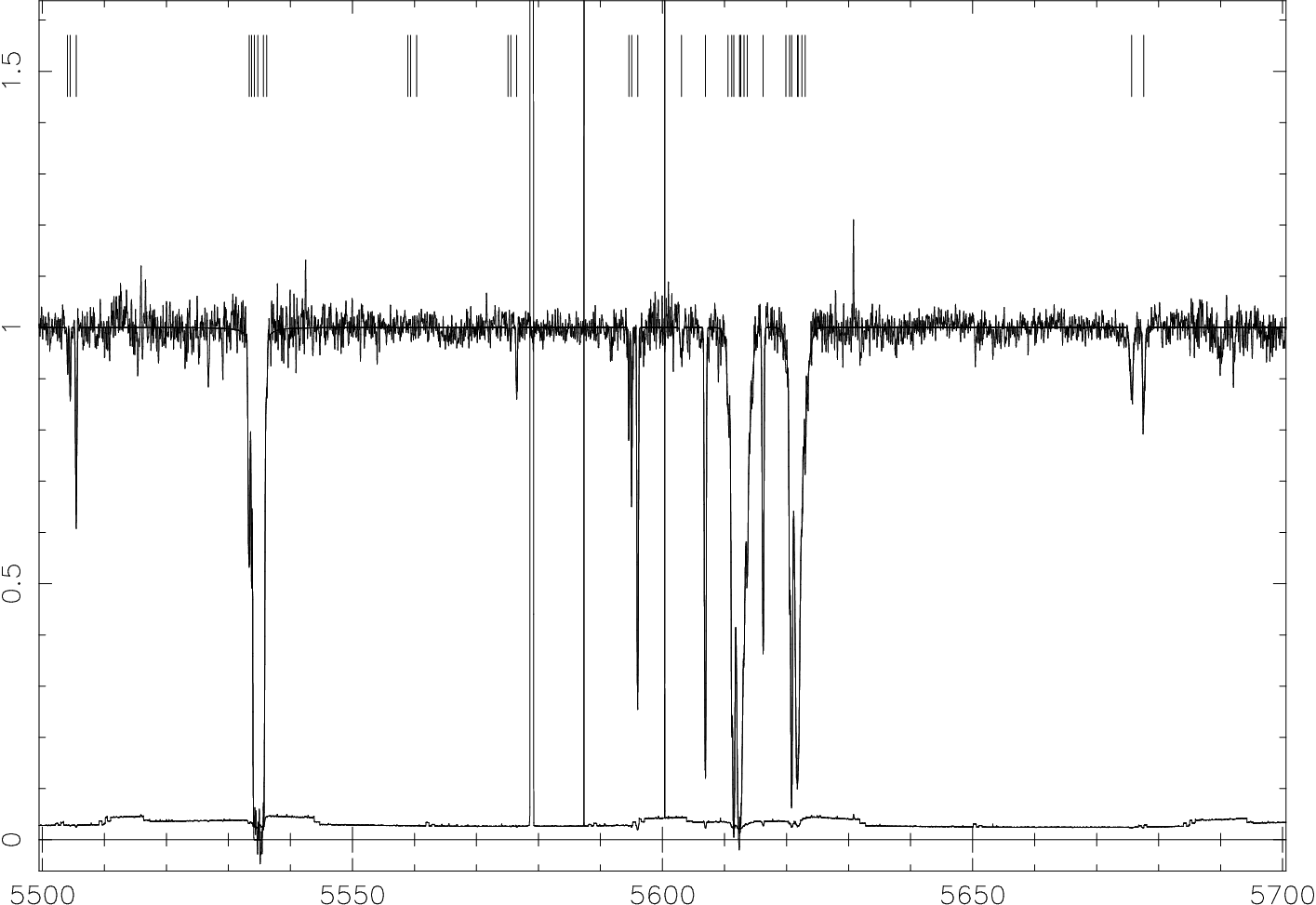,height=5.5cm,width=16cm,angle=270}}}
\caption{The spectrum of GB1759+7539, plotted against vacuum
heliocentric wavelength(\AA), normalized to unit continuum, together
with overlying profile fits. The tick marks show the position of the
components given in Table 4. The 1$\sigma$ error is also shown.}
\end{minipage}
\end{figure*}

\setcounter{figure}{2}
\begin{figure*}
\begin{minipage}{170mm}
\centerline{\hbox{\psfig{figure=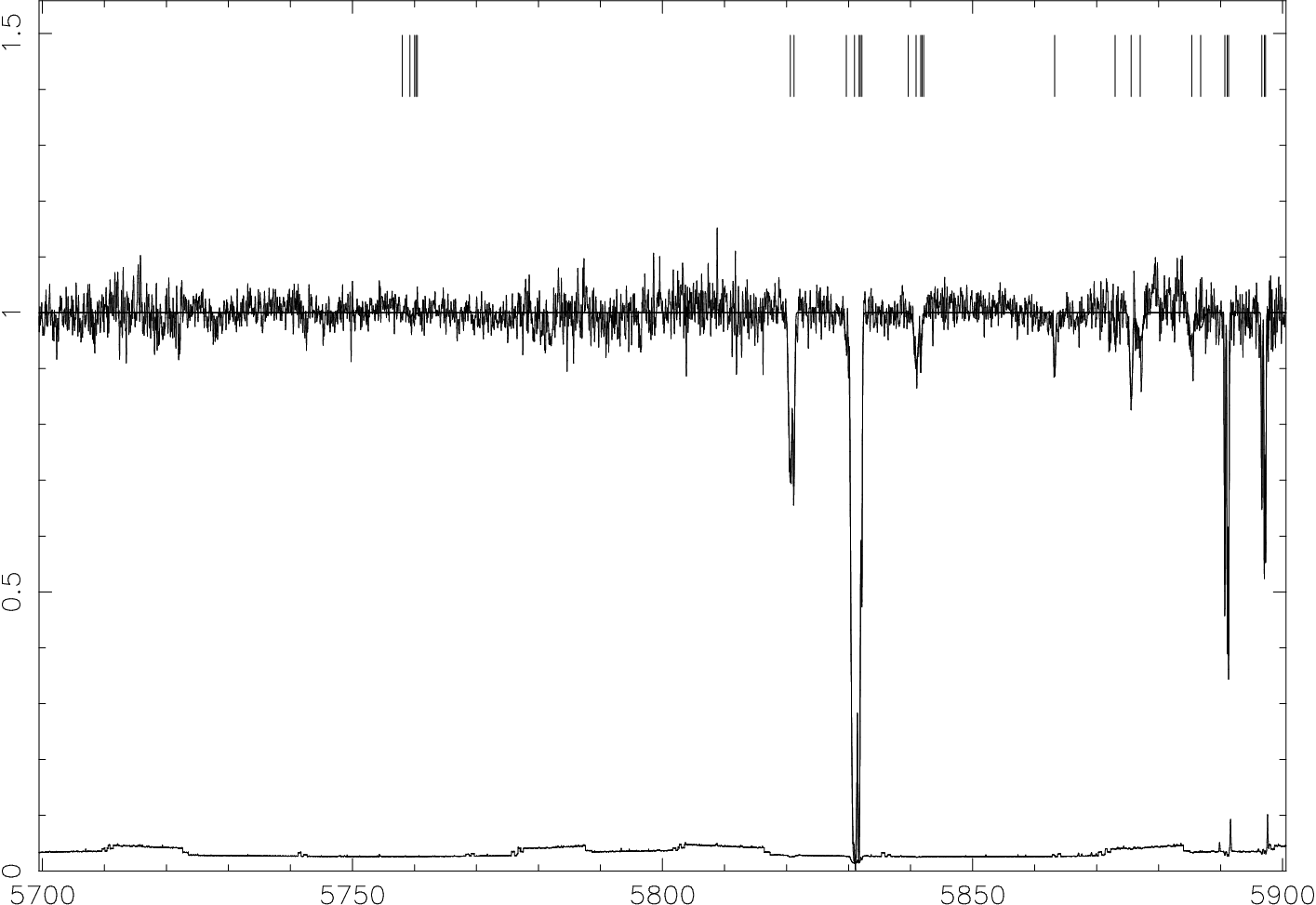,height=5.5cm,width=16cm,angle=270}}}
\centerline{\hbox{\psfig{figure=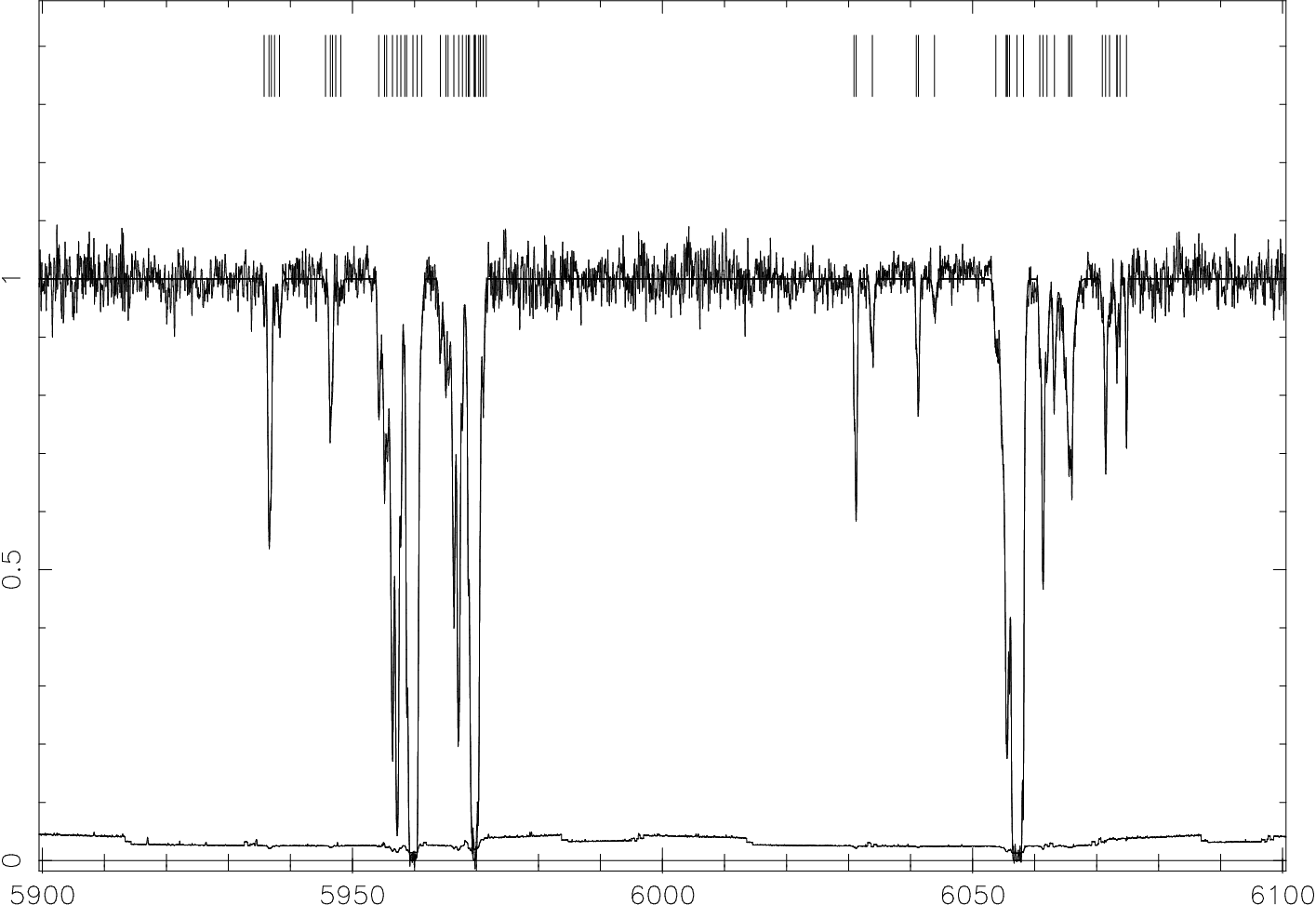,height=5.5cm,width=16cm,angle=270}}}
\centerline{\hbox{\psfig{figure=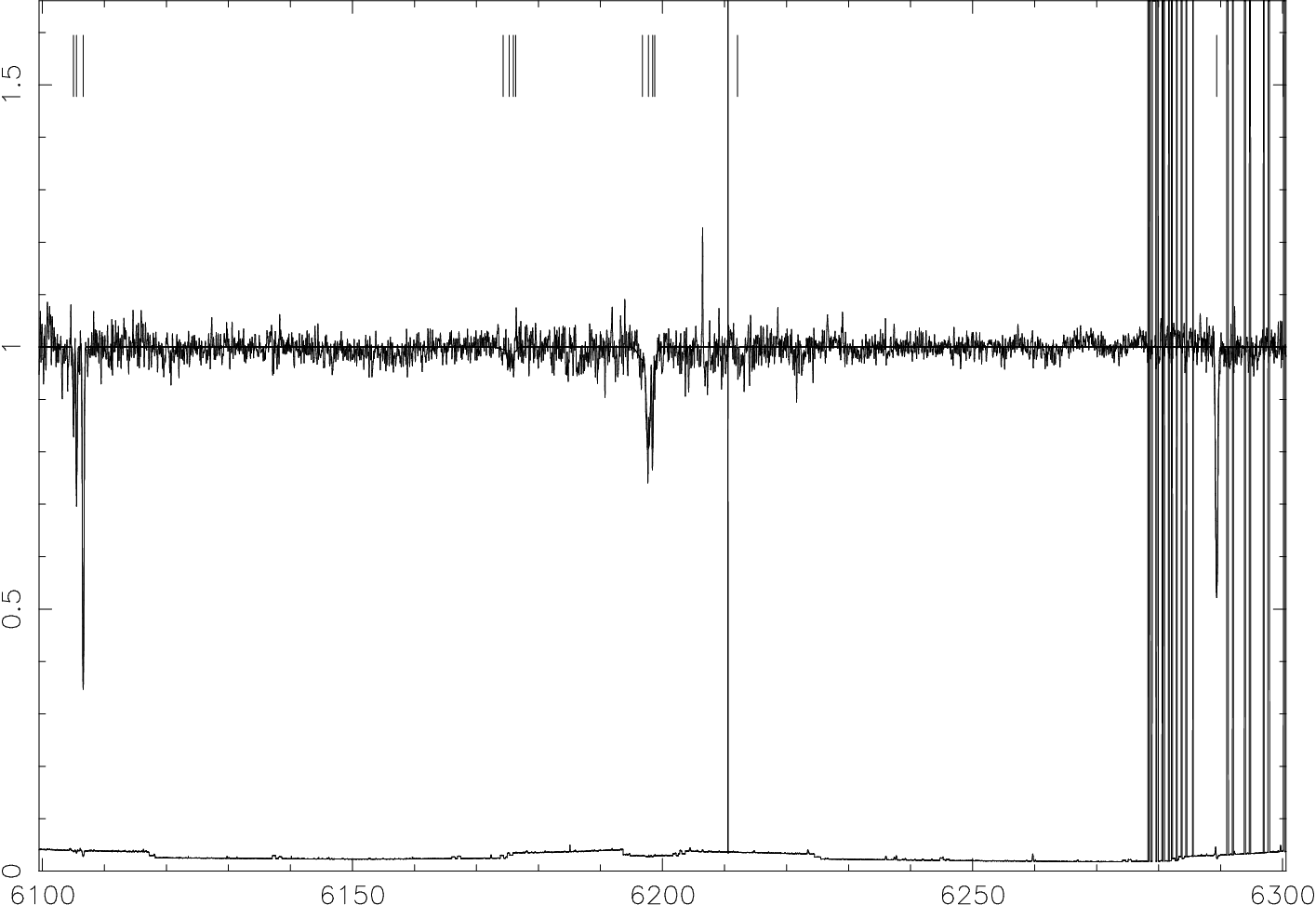,height=5.5cm,width=16cm,angle=270}}}
\centerline{\hbox{\psfig{figure=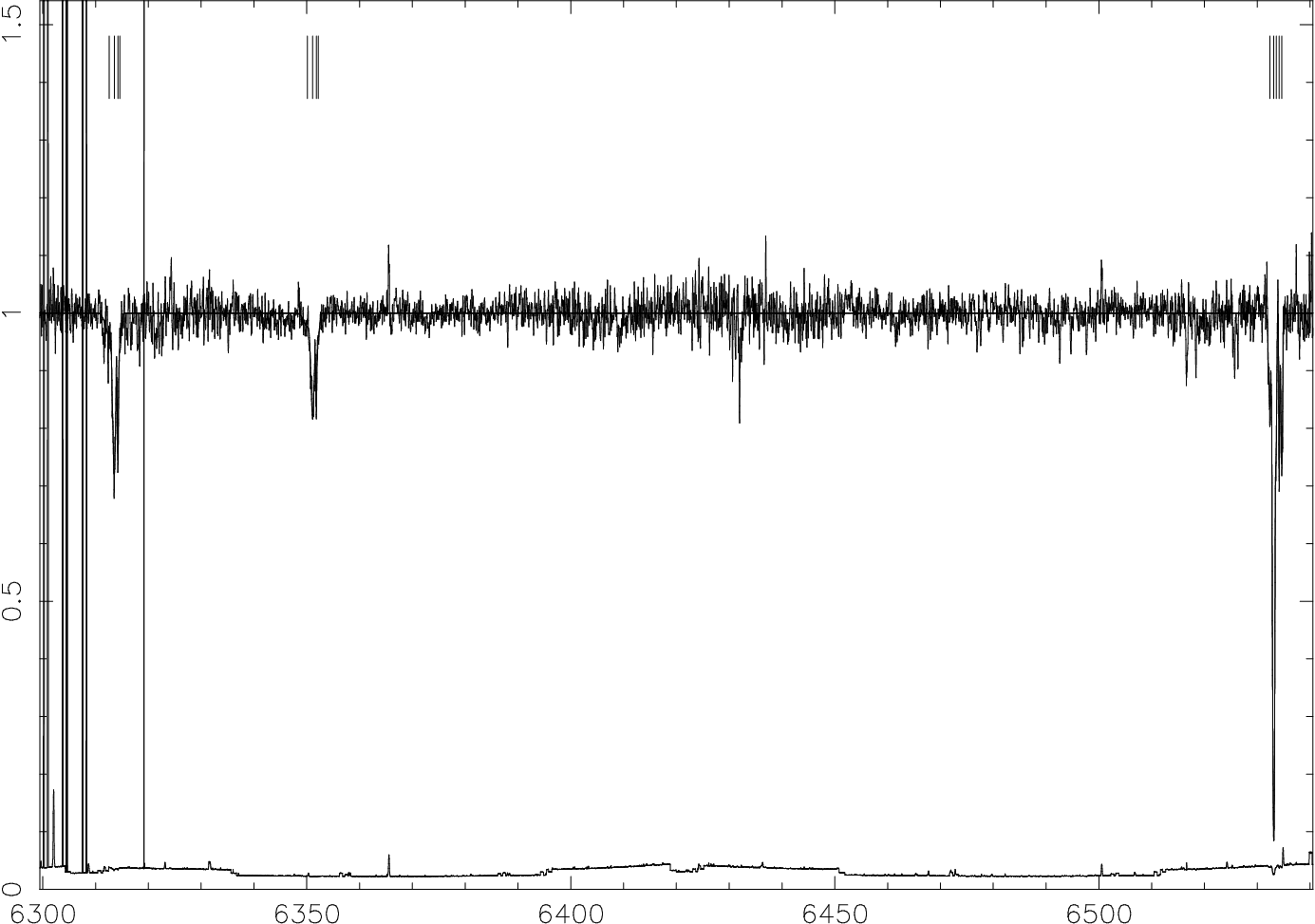,height=5.5cm,width=16cm,angle=270}}}
\caption{The spectrum of GB1759+7539, plotted against vacuum
heliocentric wavelength(\AA), normalized to unit continuum, together
with overlying profile fits. The tick marks show the position of the
components given in Table 4. The 1$\sigma$ error is also shown.}
\end{minipage}
\end{figure*}

\section{The Lyman-alpha Forest}

With one line in three positively identified as a heavy element line within
the Ly$\alpha$\  forest region of the spectrum, the risk of
contamination from further unidentified heavy element lines in the sample of
H$\:${\small I} lines appears large. Heavy element lines fitted as hydrogen
tend to have very small Doppler parameters and large column density
errors. 

The Ly$\alpha$\  lines were fitted without the additional constraint
of higher-order lines, as the observed spectrum only covered a small
fraction of the Lyman-$\beta$ region. For some features, especially
where Ly$\alpha$\  clouds are blended with saturated heavy element features,
the Voigt profile fit can be ill-constrained, and lead to H$\:${\small I}
column densities with large errors. We attempted to minimize contamination, and remove ill-constrained lines using an estimated error cutoff of $\sigma_{\rm N(H\:I)}=0.5$ dex. Only two such H$\:${\small I} lines appear, and both were simultaneously rejected using other criteria.

Incompleteness is a problem at the low column density end of the
 distribution.  Line blending, where two or more lines cannot be
 individually resolved and hence are fitted by a single Voigt profile,
 is likely to give an over-density of broad lines and a paucity of low
 column density lines in the observed distribution relative to the
 intrinsic distribution. The probability of confusion due to cloud
 blending is high because of the large number of low density
 clouds. This is especially apparent for lines with $N$(H$\:${\small
 I})$<10^{13}$cm$^{-2}$. Some weak lines, especially broad ones, may
 also be missed simply due to the finite S/N, and uncertainty in the
 continuum level. In the fitting process, the maximum Doppler
 parameter allowed was 100 km$\,$s$^{-1}$, because very few lines larger than this are expected, and they would tend to be very poorly constrained, and could be artifacts of the fitted continuum level.

The effect of line blanketing, where weak lines are lost in the
absorption profile of a stronger line, will also lead to an
under-density of weak lines.  The sample contains a few low-b lines
with low column densities, $N$(H$\:${\small I})$<10^{13}$cm$^{-2}$, that
are more likely to be fitted noise features than the intrinsic
distribution \cite{rau92}.

To avoid any
adverse effects from the proximity effect, all Ly$\alpha$\  clouds 
within 8 comoving Mpc of the QSO (38 in total) were left out. Only those lines with $13.5 <$log$N < 14.5$, were used to obtain a complete sample with reliable column density determinations.  Finally, an error cutoff of  $\sigma_{\rm N(H\:I)}<0.5$ dex, and $\sigma_b <
Max[10,10\sqrt{b/20}]$ was used, eliminating a further 2 systems, in order to minimise the effects outlined above. After all these
restrictions were taken into account, our final sample consisted of 66
lines.

The column density and Doppler parameter distributions are in good agreement with those derived by Hu et al. (1995), Lu et al (1996b) and Kirkman \& Tytler (1997). Over the limited range in column density available ($13.5 <$log$N < 14.5$), the column density distribution is
consistent with a single power-law with index $\beta = 1.68 \pm 0.15$. 
The Doppler parameter distribution peaks at around $b=23$ km$\,$s$^{-1}$, with a cut-off below about $b=18$ km$\,$s$^{-1}$, and a large tail towards high $b$-values. It is best fit by a Gaussian of mean $b=26$ km$\,$s$^{-1}$, and standard deviation $\sigma=12$ km$\,$s$^{-1}$ with a cut-off at $b=20$ km$\,$s$^{-1}$. No correlation was seen between the column density and Doppler width of the lines.

\subsection{Clustering properties}
\begin{figure}
\centerline{\hbox{\psfig{figure=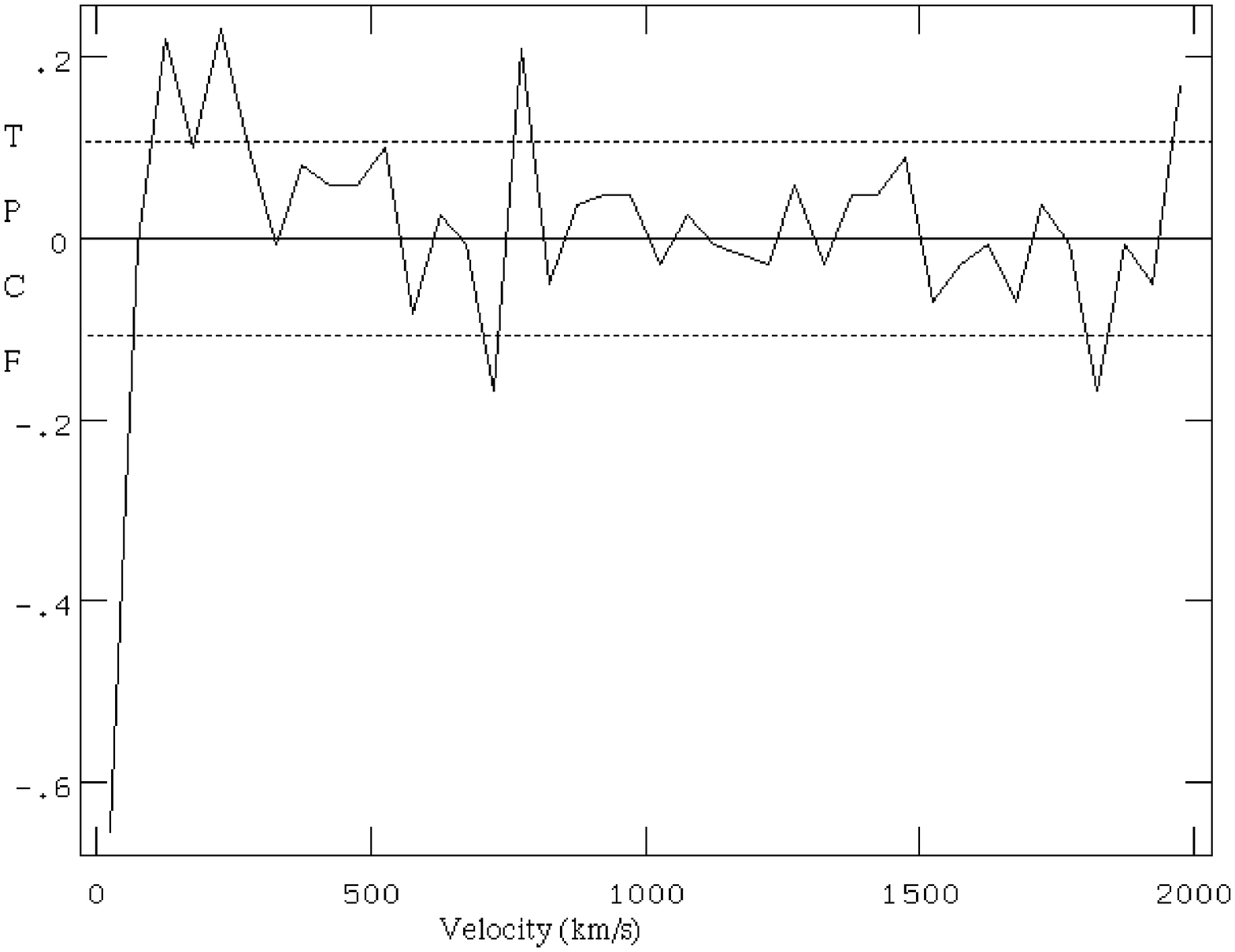,height=6cm}}}
\caption{The two point correlation function vs. velocity separation
for the entire sample of lines. The dashed lines show the 1$\sigma$
errors.}
\label{fig5}
\end{figure}
To investigate the clustering properties of the forest lines we
calculated the two point correlation function (TPCF)
\begin{equation}
\xi(v)=\frac{N_{\rm obs}(v)}{N_{\rm exp}(v)}-1
\end{equation}
where $N_{\rm obs}(v)$ is the number of observed pairs at separation $v$,
and $N_{\rm exp}(v)$ is the number of expected pairs if the lines were
randomly distributed along the line of sight. $\xi(v)$ is shown in
figure~\ref{fig5}. The TPCF shows a deficit of 61 lines pairs with separations below
50 km$\,$s$^{-1}$. As the most likely $b$-value was found to be about
25 km$\,$s$^{-1}$, and a significant number of lines had Doppler widths
considerably larger than this, the deficit could entirely be due to
line blanketing, where one line obscures another, or line blending,
where two lines would be fitted as one. These blended lines could also
help explain the non-Gaussian tail in the $b$-distribution. This effect
could also lower the observed value of $\xi(v)$ for 50$<\Delta
v<$100 km$\,$s$^{-1}$, and so we shall focus on correlations on scales
larger than this.

There is only marginal evidence of clustering along the line of sight
over the velocity range 100$<\Delta v<$250 km$\,$s$^{-1}$. We find
$<\xi>$=0.18$\pm$0.06 over this range. There is no evidence for any
clustering on scales larger than this. This result is similar to that
of Hu et al. (1995) who found 0.17$\pm$0.045 over the range
50$<\Delta v<$150 km$\,$s$^{-1}$. Lu et al.  (1996b) also noted a
weak, statistically marginal, clustering signal over a similar
velocity range. Cristiani et al.  (1997) found a small but
significant signal at $\Delta v=$100 km$\,$s$^{-1}$ of
$<\xi>$=0.2$\pm$0.04. Further analysis showed that the signal was due
to strong clustering in the larger Ly$\alpha$\  clouds, and that lines
with $N$(H$\:${\small I})$<10^{13.6}$cm$^{-2}$ showed no evidence for
clustering. Crotts (1989), Chernomordik (1995), and Hu et al. 
(1995) have also noted a stronger correlation with larger
clouds. Kirkman \& Tytler (1997), however, found no signal on any
scale, with  $<\xi>$=0.06$\pm$0.045 in the range 50$<\Delta
v<$150 km$\,$s$^{-1}$.

Cen et al. (1997) analysed Ly$\alpha$\  clouds in simulated
spectra, using a $\Lambda$CDM model, and predicted a significant
positive correlation of $<\xi>$=0.1 - 1.0 on separations of 50$<\Delta
v<$300 km$\,$s$^{-1}$ at z=3. Due to the finite box size, this is in fact
an underestimate of the correlation in these simulations. Our results
are in line with the lower end of the prediction, but rule out
clustering on a scale $<\xi>\sim$1.0.

\subsection{The 1-point and 2-point joint probability distribution}
\begin{figure}
\centerline{\hbox{\psfig{figure=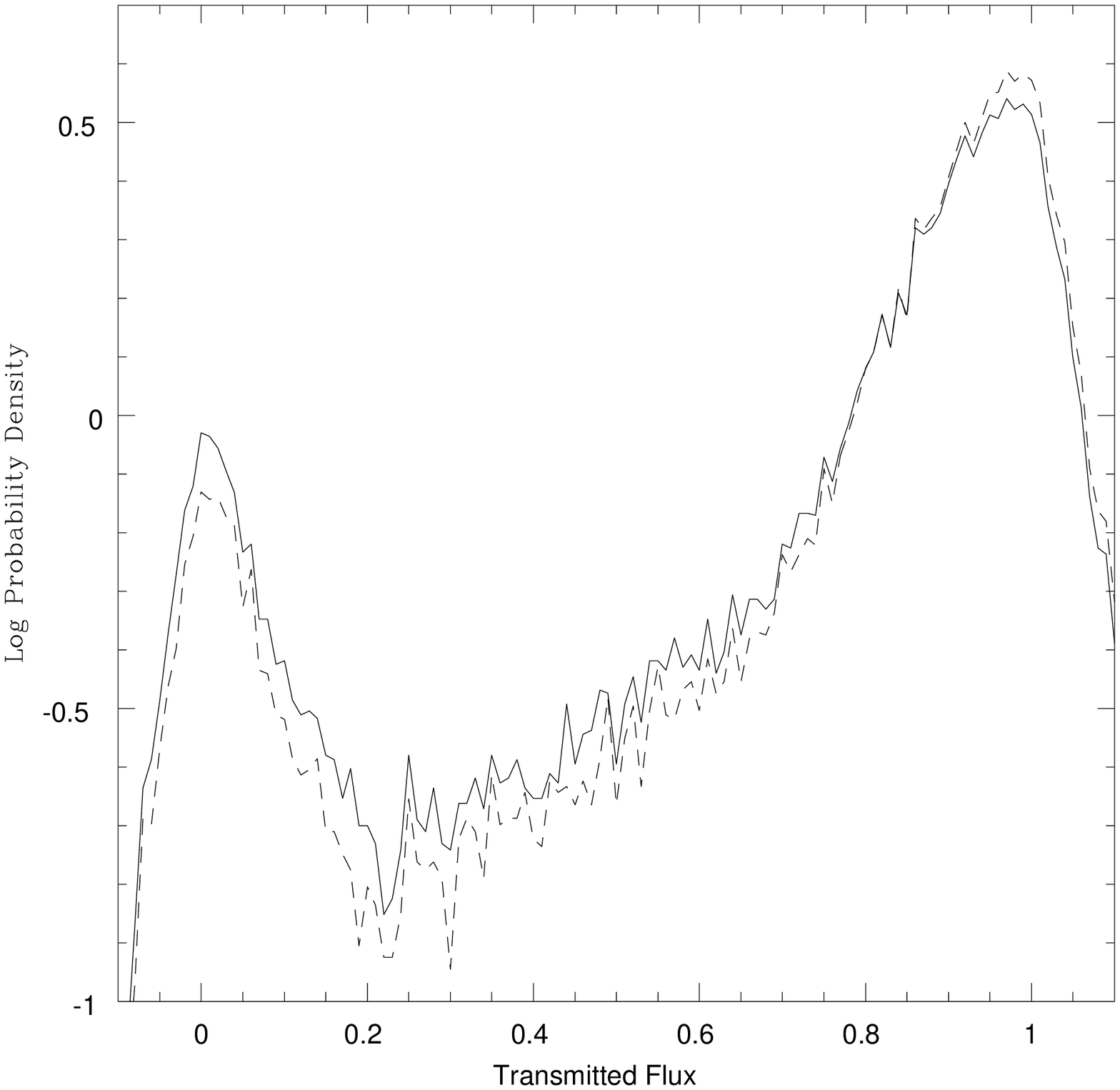,height=6cm}}}
\caption{The flux distribution at $<z>$=2.7. The solid line is the
result for the whole Ly$\alpha$\  forest spectrum, except for the
damped systems which were removed. For the dashed line, all of the
identified heavy element lines were removed as well.}
\label{fig6}
\end{figure}

The recent N-body cosmological simulations of the Ly$\alpha$\  forest
(e.g. Cen et al. 1994; Miralda-Escud\'e et al. 1996) have called into question the physical
meaning behind the Voigt profile fitting procedure. In the past,
Ly$\alpha$\  clouds were viewed as discrete clouds confined by
pressure or gravity within the intergalactic medium (IGM). If these
clouds had a Gaussian velocity dispersion then the Voigt profile would
describe the resulting line profile, and the physical column density
and Doppler width parameters could be estimated. The new paradigm,
brought about by the cosmological simulations, views the Ly$\alpha$\
forest as fluctuations in the IGM itself. Some of the Ly$\alpha$\
forest lines in the simulations are seen to be non-Voigt, due to
gravitational infall, or flows in the IGM. In the simulation models,
the Ly$\alpha$\  clouds have a variety of shapes and sizes, from
nearly spherical clouds, to elongated filamentary structures, and
cannot be described by a simple model \cite{cs97}.

The Voigt profile analysis also has large intrinsic uncertainties due
to line blending seen in high quality, high redshift data. Different
initial guesses, and number of lines fitted to a region may determine
different local minima in the $\chi^2$ parameter space \cite{kt97}
This non-uniqueness is not helped by increasing S/N. This has lead to the
consideration of other statistics when comparing the
observed quasar absorption spectra with those from
simulations. Following Miralda-Escud\'e et al.  (1997), we have
measured the 1-point, and 2-point joint probability distribution of
the transmitted flux in the Ly$\alpha$\  forest spectrum of
GB1759+7539.

The Ly$\alpha$\  forest region of GB1759+7539 is very rich in heavy
element features, with over 30\% of the Voigt profiles fitted
positively identified as heavy element lines. Heavy element absorption lines
tend to be much narrower than Ly$\alpha$\  lines, due to the lower
thermal velocity dispersion of the heavier ions. We have investigated
the effect that heavy element contamination has on the flux statistics, to
test their discriminatory power.

\begin{figure}
\centerline{\hbox{\psfig{figure=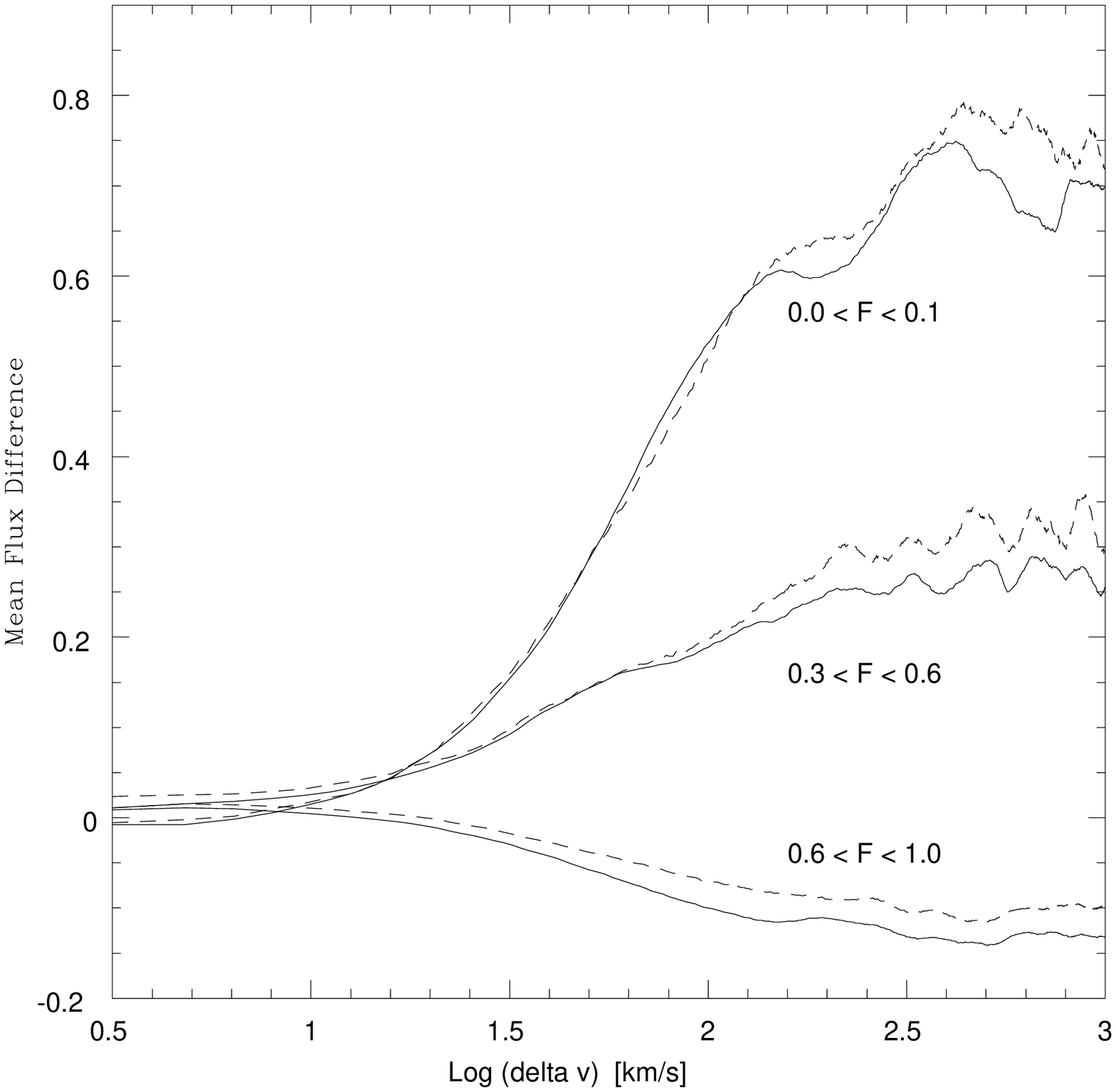,height=6cm}}}
\caption{The mean flux difference distribution at $<z>$=2.7, averaged
over the flux intervals F as indicated in the figure. The solid lines
are the results for the whole Ly$\alpha$\  forest spectrum, except for
the damped systems which were removed. For the dashed lines, all of
the identified heavy element lines were removed as well.}
\label{fig7}
\end{figure}

The 1-point and 2-point joint probability distribution of the
transmitted flux were calculated on the whole available spectrum,
excluding the regions where the damping wings of the two large
systems suppressed the spectrum. The results are shown as the solid
lines in figures~\ref{fig6} and ~\ref{fig7}. Then, in order to
investigate the effect of the heavy element features, all of the identified
heavy element lines were masked off, and the distributions were
calculated again. The results for the heavy element-free regions of the
spectrum are shown as the dashed lines in figures~\ref{fig6} and
~\ref{fig7}.

It is clear, from studying figure~\ref{fig6}, that the effect of
masking out the heavy element lines is to slightly raise the average
transmitted flux. This is to be expected, as only regions where heavy element
absorption was present were masked off, and not regions where no
absorption was seen. The shape of the flux probability distribution,
however, appears almost unchanged.

The effect of raising the average transmitted flux when the heavy element
lines are masked off also explains the higher levels of the mean flux
difference at high velocity separations in the heavy element-free spectrum
seen in figure~\ref{fig7}. The shape of the 2-point joint probability
distribution at separation velocities $\delta v < 100$ km$\,$s$^{-1}$ is a
good indication of the mean profile of the absorption lines. For $0.0
< F < 0.1$, the curve depends on the mean shape of saturated lines,
whereas the $0.6 < F < 1.0$ curve is sensitive to weak absorbers. It
can be seen, however, that there is virtually no difference between
the two samples, with and without the heavy element lines. This indicates that
the statistic does not detect heavy element contamination, or equivalently, it is not a good discriminator between samples where a large minority of
lines are drawn from a different, intrinsically narrower distribution,
and samples where no such narrow lines are present. 

\subsection{Departures from the Voigt profile}

A clear signature of deviations from Voigt profiles is observed in the
simulated forest spectra \cite{mir96}, typically in systems still in
a collapse phase. The cool, dense gas in its core produces a high
column-density, narrow component, whereas the bulk motion, and
shock-heating of the infalling material typically produces a broader
component \cite{rau96}, producing departures, often asymmetric, from
the Voigt profile. Although Voigt profile fitting yields an excellent
fit to the data, it is expected that information about this intrinsic
non-Voigtness is contained in the way that small, or broad lines
cluster about intermediate strength Ly$\alpha$\  lines.

In order to investigate this, we decided to remove all of the strong
Ly$\alpha$\  lines from the spectrum, leaving just the small clouds
that may show some such evidence of clustering. This was achieved by
dividing the spectrum of GB1759+7539 by the Voigt profile fits of all
heavy element lines, and all the hydrogen lines of Ly$\alpha$\  forest clouds
with a column density $N$(H$\:${\small I})$>10^{13}$cm$^{-2}$. All
points of the spectrum with a flux level of less than 0.2 times the
continuum level were not included in the division, and were then given
zero weighting in the following addition. This spectrum should
therefore still contain the signature of the small infalling clouds,
which would tend to be fitted with Voigt profiles of column density
$N$(H$\:${\small I})$<10^{13}$cm$^{-2}$, with the central narrow
components now removed. Although this signature would be difficult to
see in any single system, due to the low S/N and blending effects, it
can be searched for statistically by stacking many such systems with
the central Ly$\alpha$\  component removed. If the departures from a
Voigt profile were significant, one would expect to see a slight dip
in the residual flux either side of the central wavelength in this
composite spectrum.

\begin{figure}
\centerline{\hbox{\psfig{figure=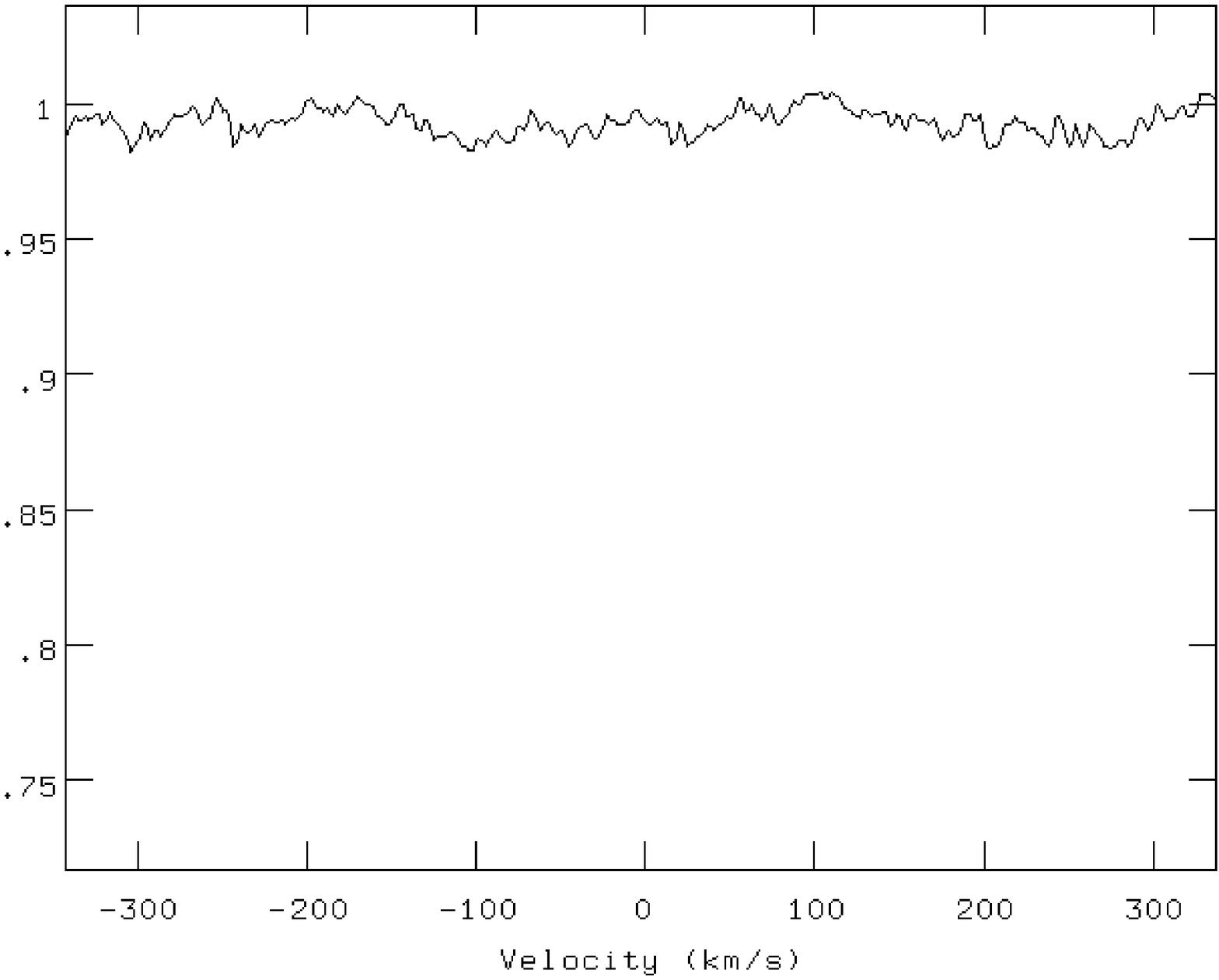,height=6cm}}}
\caption{Composite spectrum for 150 H I lines with column densities in
the range $10^{13}<N$(H$\:${\small I})$<10^{14}$cm$^{-2}$, after
division by the Voigt profile fit, as described in the text. The
spectrum is shown in velocity space relative to H I
$\lambda1215.67$\AA.}
\label{fig10}
\end{figure}

A composite spectrum was formed by shifting the divided spectrum of
each system to the rest frame of the removed Ly$\alpha$\  line,
rebinning onto a common velocity scale, and averaging all such
rest-frame spectra, using variance weighting. 150 intermediate
strength ($10^{13}<N$(H$\:${\small I})$<10^{14}$cm$^{-2}$) Ly$\alpha$\
lines were used, giving a final S/N $\sim$ 400. The composite spectrum
can be seen in figure~\ref{fig10}.

There is no significant evidence in the composite spectrum (figure~\ref{fig10}) of
Voigt profile departures due to infalling gas, as described in
Miralda-Escud\'e et al. . The spectrum shows no significant dip
either side of the central wavelength, and is consistent with the
smaller Ly$\alpha$\  lines being distributed randomly along the
line-of-sight. Different column density limits, both for the
summation, and for inclusion in the division were also used, but the
results were similarly featureless. Similar tests need to be performed on simulated data to see if this constraint is an important one.

\section{The Heavy Element Systems}
In the line of sight to GB1759+7539 there are two large systems with conspicuous damping wings, as
well as numerous smaller heavy element systems. One of the two systems has a column density $N$(H$\:${\small I})$> 2\times 10^{20}$; the established, yet somewhat arbitrary criterion for a damped Lyman-alpha system (DLA) \cite{wol86}. The other system has a slightly lower column density ($N$(H$\:${\small I})$= 6\times 10^{19}$), but displays similar properties to the larger DLAs. In this section we will study
the elemental abundances of these two systems. We will only
consider those abundance measurements that appear free from saturation
effects, and use Fe as the metallicity indicator. The effects of dust
on the measured abundances, the ratios of elements produced in
different nucleosynthesis processes, and the implications for galactic
chemical evolution will be discussed.

\subsection{Dust depletion}

In the diffuse inter-stellar medium (ISM) clouds, the relative intrinsic 
abundances are
believed to be solar, and so any departures are thought to be due to
varying levels of dust depletion \cite{ss96}. The refractory
elements, with a high condensation temperature, are heavily depleted
by the formation of dust grains, whereas the abundance of elements
with a low condensation temperature is largely unaffected. Generally,
the elements S, P, Zn, C, N, and O are depleted by a factor about 3 or
less, but elements Si, Fe, Cr, Al, and Ni are much more heavily
depleted in the ISM.

High redshift, young galaxies may have a different chemical enrichment
pattern than present day galaxies, due to the different timescales of
the various nucleosynthesis processes. In order to objectively study
the dust depletion seen in DLAs,  one
therefore needs to compare the abundances of elements produced in the
same nucleosynthesis processes that have very different depletion
patterns. Vladilo (1998) studied the abundances of Zn, Fe, and Cr in a
sample of 17 DLAs, and concluded that the dust-to-gas ratio was
between 2 and 25 \% of the Galactic value. This is in good agreement
with the value of 5 - 20 \% estimated by Pei, Fall \& Bechtold (1991)
from their study of the reddening of QSOs with foreground DLA
absorption. Pettini et al.  (1997a) concluded that the
``typical'' dust-to-gas ratio of DLAs is only $\sim1/30$ of that of
the Milky Way. Lu et al.  (1996a), however, assert that there
is no significant evidence for any dust depletion and that the
overabundance of Zn relative to Cr may be intrinsic.

\subsection{Nucleosynthesis and abundance ratios}

Detailed abundance analysis of the disk and halo stars has provided
evidence of the chemical evolution of our Galaxy \cite{mcw97} The
chemical composition of low mass stars has changed little since their
formation, and so by studying the abundance ratios in old, low
metallicity stars, we can learn about the nucleosynthesis processes
that took place when our Galaxy was forming.

The halo stars have an average metallicity of [Fe/H]$\sim-1.6$. They
have enhanced [$\alpha$/Fe] ratios (where $\alpha$ includes O, Mg, Si,
and S) by a factor of about 3 relative to solar \cite{mcw97} due to
the $\sim 10^8$ year time delay between $\alpha$-element producing
SNII and the first SNIa; the main source of Fe-peak elements
\cite{tin79}. This over-abundance has also been observed in DLAs
\cite{lu96a,max95}.

The production of odd atomic number elements depends on the neutron
excess, which in turn depends on the initial metallicity of the
nuclear fuel \cite{ta71}. This leads to an under-abundance of odd
elements, relative to the even atomic number $\alpha$-elements
(e.g. [Al/Si]) at low metallicity. The [Al/Fe] ratio, however,
increases slightly with decreasing metallicity in Galactic stars
\cite{mcw97} indicating that Al could be classified as a mild
$\alpha$-element phenomenologically despite having an odd number of
protons.

Some chemical evolution models also predict that nitrogen may be
underabundant relative to oxygen in young high redshift objects, due
to the delayed release of primary nitrogen in intermediate mass stars
relative to oxygen from high mass stars \cite{vil93}. This effect, 
however, has not
been convincingly seen in H$\:${\small II} regions of nearby
heavy element-poor galaxies \cite{thu95}, or in blue compact galaxies \cite{izo}. It has been investigated in high
redshift DLAs by Pettini et al.  (1995) and Lipman et al. (1995). Later Lu et al.  (1998) found a large
scatter in the N/Si ratio, so giving support for the time-delay model of
primary N production. 

\begin{figure}
\centerline{\hbox{\psfig{figure=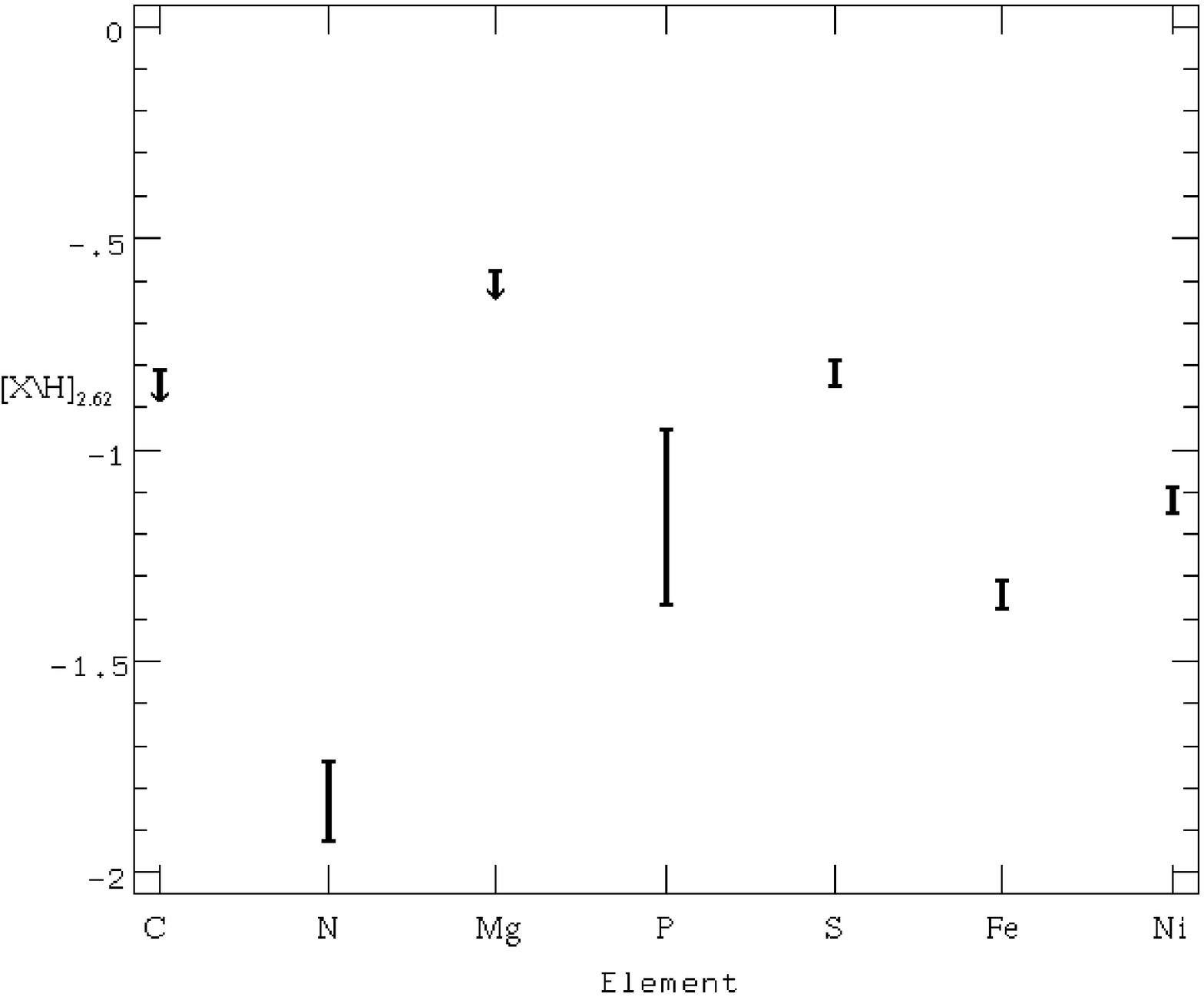,height=6cm}}}
\caption{Observed elemental abundances relative to solar for the
$z_{\rm abs}=2.625$ DLA plotted on a logarithmic scale. The errors, and
limits are intrinsic and given to 3$\sigma$. The upper limit for C$\:${\small II}
is derived by considering the abundance of C$\:${\small II}* and the temperature of
the cosmic microwave background at that epoch.}
\label{fig8}
\end{figure}

\subsection{The Damped Ly$\alpha$\  System at $z_{\rm abs}=2.625$}
The measured abundances for the $z_{\rm abs}=2.625$ DLA are listed in
table 2, and plotted in figure~\ref{fig8}. The velocity profiles of many of the absorption features can be seen in figure~\ref{figv1}. In calculating the
abundances, it was assumed that the ion species observed were from the
dominant ionization stages in H$\:${\small I} gas, so that corrections
for other ionization stages are negligible. Given the large
H$\:${\small I} column density of the system, this is believed to be a
good assumption \cite{vie95}. Also, it is assumed that the
H$\:${\small I} column density is a good measure of the hydrogen
content of the system, since the fraction of molecular hydrogen is
known to be small \cite{lev92}.

Another source of uncertainty is that, as seen in the velocity profile
of the heavy element lines, there are several components contributing to this
system. In the H$\:${\small I} feature, however, this structure is not seen, and it
is possible that there may be a metallicity gradient, or additional
components contributing to the Ly$\alpha$\  line, but not the heavy element
lines. We summed the column densities for the heavy element lines over all the
components, and so are examining the average properties of the
DLA. Although the metallicity may vary through the object, it appears,
by comparing the column density of individual heavy element line components,
that the relative abundances are fairly constant. The solar abundance
values used in the calculation were taken from Anders \& Grevesse
(1989).

From the abundance of iron, [Fe/H]$=-1.34\pm0.01$, calculated using
the unsaturated $\lambda=1611$   line, we observe that the
$z_{\rm abs}=2.625$ DLA has a metallicity $Z_{\rm 2.62}\simeq1/20 Z_{\rm \odot}$,
which is about typical for a DLA at this redshift (Pettini et al. 1997b). Hence, this galaxy is still in the early stages of its
chemical evolution. Note, however, that in reaching this conclusion,
it is assumed that there is little depletion of Fe onto dust
grains. If much dust is present, the system would have a higher
metallicity than estimated.

\subsection{The System at $z_{\rm abs}=2.91$}
\begin{figure}
\centerline{\hbox{\psfig{figure=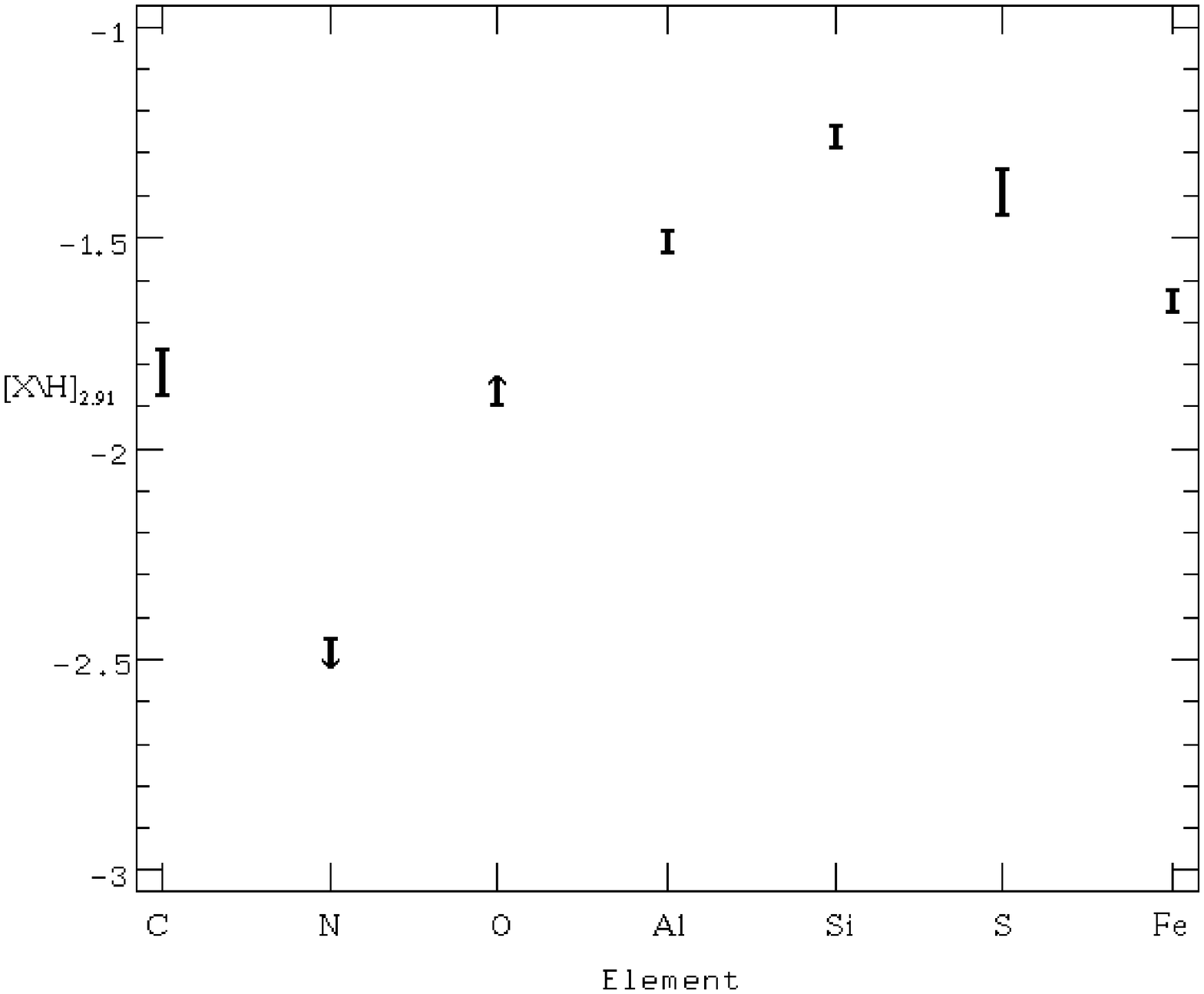,height=6cm}}}
\caption{Observed elemental abundances relative to solar for the
$z_{\rm abs}=2.911$ plotted on a logarithmic scale. The errors, and
limits are intrinsic and given to 3$\sigma$.}
\label{fig9}
\end{figure}

The heavy element system at $z_{\rm abs}=2.91$ has a hydrogen column density of
Log N(H$\:${\small I})=19.8. The heavy element
abundances were calculated as described above, assuming no ionization
correction. Due to the smaller size, and lower metallicity of this
system, different ion species were on the linear regime of the curve
of growth, and hence the calculated abundances are for different
species than the $z_{\rm abs}=2.625$ system. Only the central component of the weak Fe$\:${\small II} feature is seen (see figure~\ref{figv2}), so to produce  fair relative abundance estimates, only the corresponding components of the other elements were considered. Ignoring the weaker outlying components may lead to an underestimate of the absolute abundances, and hence the overall metallicity of the system, by about 0.1 dex. The measured abundances are
listed in table 3, and plotted in figure~\ref{fig9}. Assuming little
dust depletion, the metallicity of this system is
$Z_{\rm 2.91}\simeq1/45 Z_{\rm \odot}$. The velocity profiles of many of the absorption features can be seen in figure~\ref{figv2}.

\subsubsection{Deuterium Abundance Determination?}

Burles \& Tytler \shortcite{bt98} attempted to determine the cosmologically important D/H ratio by observing the Ly$\alpha$, Ly$\beta$, and Ly$\gamma$\ absorption line profiles of this system. They claim to have measured a value of Log(D/H)$=-4.3\pm0.3$. On the basis of the evidence presented here it is hard to see how this measurement could be reliable. Due to the high column density (Log $N$(H$\:${\small I})=19.8) of the system, the hydrogen Ly$\alpha$\  line (see figure~\ref{figv2}), and to some extent the Ly$\beta$\  line, obscure the deuterium counterpart at $-80$ km$\,$s$^{-1}$. Several more Lyman series lines would be required to determine the intrinsic structure in the hydrogen absorption of the system, and distinguish this from any deuterium absorption, and any intervening Ly$\alpha$\  features. The structure seen in the heavy element absorption lines of the system (7 components of Si$\:${\small II} were fitted, spanning over 150 km$\,$s$^{-1}$; see figure~\ref{figv2}) suggests that even then an accurate determination may not be possible. Unfortunately, they published only the result, and not the full analysis.

\subsection{Discussion: Dust and the intrinsic abundances}

The over-abundances of sulphur relative to iron in the two
systems, [S/Fe]$_{\rm 2.62}=0.5$, and [S/Fe]$_{\rm 2.91}=0.3$ and silicon in the $z_{\rm abs}=2.91$
system, [Si/Fe]$_{\rm 2.91}=0.4$ are consistent with that seen in
$\alpha$-elements at this metallicity \cite{mcw97}.  As sulphur is
undepleted by the formation of dust grains, yet silicon is heavily depleted, 
the measurements for
the $z_{\rm abs}=2.91$ system imply that there must be a low level of dust
depletion in this system. 

The relative abundance of carbon, [C/Fe]$_{2.91}= -0.17$, in the
smaller system has an essentially solar value. Wheeler et al. (1989)
concluded that the abundance of carbon is essentially solar,
irrespective of metallicity. A slight enhancement of C was observed
in the Galactic disk at low metallicity \cite{tom95}, but this was
not seen in halo stars \cite{mcw95}. The important point is that carbon 
has not been seen to be {\it under}-abundant at any
metallicity. Since carbon is relatively undepleted by dust, but iron
is, any dust depletion in this system would imply an intrinsic
under-abundance of carbon, and so this adds further weight to the 
conclusion that the level of dust
in the $z_{abs}=2.91$ system must be very low.

Prochaska \& Wolfe (1999) also observed GB1759+7539 using HIRES on the Keck telescope, with a higher wavelength setting, and concentrated on the absorption features of the $z_{\rm abs}=2.62$ system as part of a survey of the chemical abundances of several DLAs. They measured the column density of Fe$\:${\small II}, also using the unsaturated $\lambda=1611$  line, and obtained an abundance of [Fe/H]$_{\rm 2.62}=-1.23\pm0.1$, higher than our measurement by 0.1 dex. Their coverage did not include the Ly$\alpha$\  forest, and so a less accurate H$\:${\small I} column density estimate was used, explaining the larger quoted errors. Ni$\:${\small II} was also observed. An abundance of [Ni/H]$_{\rm 2.62}=-1.48\pm0.1$ was obtained using older oscillator strength values. This is 0.1 dex lower than our determination, after correction for the f-values used. The differences are fairly small, and probably due to different model inputs (e.g. the number of Voigt profile components). Their observations are therefore entirely consistent with ours. 

Their most interesting measurement is that of Si$\:${\small II}, using the unsaturated $\lambda=1808$ line. They find [Si/H]$_{\rm 2.62}=-0.82\pm0.1$; exactly the same as our measurement for [S/H]$_{\rm 2.62}$. These two elements are both $\alpha$-elements, produced in the same nucleosynthesis processes, and yet have very different dust depletion patterns. The fact that they have the same relative abundance therefore implies that this system has a very low dust content. Cr$\:${\small II}$\lambda 2066$ was observed with an abundance of [Cr/H]$_{\rm 2.62}=-1.27\pm0.1$, entirely consistent with the Ni and Fe measurements. Unfortunately, however, the Zn$\:${\small II}$\lambda 2026$ feature was blended with sky lines, making an abundance determination uncertain.

Chemical evolution models predict that there should be a large scatter
in the values of [N/O] due to the delayed release of primary
nitrogen. The O$\:${\small I} $\lambda 1302$ line is
saturated in both systems, so a comparison with S is considered instead. This is based
on the assumption that the ratio O/S is essentially solar. We find
that [N/S]$_{\rm 2.62}= -1.0$ and [N/S]$_{\rm 2.91}< -1.06$, which is lower than that of the heavy element-poor
H$\:${\small II} regions, and consistent with the time-delay model of primary
nitrogen production.

The abundance of phosphorus was measured at [P/Fe]$_{\rm 2.62}=0.2$. In
the smaller system, aluminium was observed, with
[Al/Fe]$_{\rm 2.91}=0.1$. P and Al have an odd number of protons, and so
should be under-abundant relative to even atomic number
elements. Indeed, [P/S]$_{\rm 2.62}= -0.3$, and [Al/Si]$_{\rm 2.91}= -0.3$ as
expected. McWilliam (1997) noted that Na, and Al are have a slight
over-abundance relative to Fe at low metallicities. Since P is formed
from Al by the addition of an alpha-particle, it too should exhibit
the same property. Our measured abundance is consistent with this.

\section{The CMB Temperature at $z=2.62$}

According to Big Bang theory, the temperature of the cosmic microwave
background (CMB) should increase linearly with $(1+z)$. The local
background temperature has been accurately measured at 2.73K. Bahcall
\& Wolf (1968) suggested that by observing the relative populations of
ground-state fine-structure lines of certain atoms, as seen in QSO
absorption systems, one could estimate the temperature of the CMB at
high redshift. In the last few years, several measurements at high
redshift have been made, using the fine-structure levels of C$\:${\small I} and C$\:${\small II}
\cite{son94a,son94b,lu96a,lu96c}, all of which were consistent with
the Big Bang prediction.

The C$\:${\small II} line of the $z_{\rm abs}=2.62$ DLA was saturated. Therefore, in
order to estimate an upper limit for the temperature of the CMB, we
assume that [C/Fe]$_{\rm 2.62}> -0.3$ and hence that log $N$(C$\:${\small II}) $>
15.7$. The ratio [C/Fe] is believed to be essentially solar at all
metallicities \cite{whe89}, and the effect of any dust would be to
increase the intrinsic metallicity, and hence the abundance of
C. Therefore, we believe this to be a fair assumption. The velocity profile for the marginally detected C$\:${\small II}* absorption feature can be seen in figure~\ref{figv1}. Although the profile appears to match that of other low ionization lines (e.g. Ni$\:${\small II}) giving confidence in the identification, it lies in the Ly$\alpha$\  forest, and hence there is the possibility of Ly$\alpha$\  contamination. Therefore, we take an upper limit (3$\sigma$) of log $N$(C$\:${\small II}*) $<13.0$. The C$\:${\small II}*/C$\:${\small II} ratio then yields an upper limit of T$_{\rm CMB} <12.9$K at $z=2.62$. This
is very close to the Big Bang cosmology prediction,  T$_{\rm CMB}=9.9$K,
at this redshift, which suggests that there is negligible excitation
of the C$\:${\small II} ions by other mechanisms.

\section{Summary and Conclusions}
We obtained an echelle spectrum of the high-redshift QSO GB1759+7539
($z_{\rm em}=3.05$) using the instrument HIRES on the Keck 10m
telescope. The spectrum has wavelength coverage 4100 - 6540\AA, a
resolution of FWHM=7 km$\,$s$^{-1}$, and a typical S/N per 2 km$\,$s$^{-1}$ pixel of $\sim 25$ in the
Ly$\alpha$\  forest region, and $\sim 60$ longward of the Ly$\alpha$\
emission. Voigt profiles were fitted to all of the absorption
features, and the heavy element lines from twelve heavy element systems were
identified. The Ly$\alpha$\  forest lines, and the absorption features
of the two large systems were analysed in detail.

The observed Ly$\alpha$\  forest systems have a mean redshift of
$<z>=2.7$.  The H$\:${\small I} column density distribution is well
described by a power law distribution with index $\beta = 1.68 \pm
0.15$ in the range $13.5 <$ log$N < 14.5$. The Doppler width
distribution is consistent with a Gaussian distribution of mean
$b=26$ km$\,$s$^{-1}$, and standard deviation $\sigma=12$ km$\,$s$^{-1}$ with a
cut-off at $b=20$ km$\,$s$^{-1}$. There is marginal evidence of clustering along the line of
sight over the velocity range 100$<\Delta v<250$ km$\,$s$^{-1}$. The
1-point and 2-point joint probability distributions of the transmitted
flux in the Ly$\alpha$\  forest region of the spectrum were
calculated, and shown to be very insensitive to the heavy element
contamination. We could find no evidence of Voigt profile departures
due to infalling gas, as observed in the simulated forest spectra
\cite{mir96}.

The  C, N, O, Al, Si, P, S, Mg, Fe, and Ni absorption features of the
large systems were studied, and the elemental abundances calculated for
the weak unsaturated lines. The systems have metallicities of
$Z_{\rm 2.62}\simeq1/20 Z_{\rm \odot}$ and $Z_{\rm 2.91}\simeq1/45
Z_{\rm \odot}$. Both systems appear to have a low dust content. They show
an over-abundance of $\alpha$-elements relative to Fe-peak elements,
and an under-abundance of odd atomic number elements relative to
even. N was found to be
under-abundant relative to O, in line with the time delay model of
primary N production.

\section*{Acknowledgements}

\noindent We would like to thank Max Pettini, Sandra Savaglio, and an anonymous referee, for helpful comments. PJO acknowledges support from PPARC. Much of the data reduction and analysis was performed on the Starlink-supported computer network at the Institute of Astronomy.

\begin{table*}
\begin{minipage}{120mm}
\caption{\rm Absorption line parameter list}
\label{tbl-4}
\begin{tabular}{lrrrrrrrr}
\hline  
ID & $\lambda_{\rm rest}$ & $\lambda_{\rm obs}$ & $z$ & $\sigma_{z}$ & $b$  & $\sigma_{b}$ & Log$N$  & $\sigma_{{\rm Log}N}$\\
\hline  
Si$\:$IV & 1402.77 & 4117.27 & 1.935102 & 0.000000 & 8.48 & 0.00 & 12.987 & 0.044 \\
Si$\:$IV & 1402.77 & 4117.80 & 1.935480 & 0.000000 & 7.93 & 0.00 & 12.407 & 0.150 \\
Si$\:$IV & 1402.77 & 4118.26 & 1.935805 & 0.000000 & 8.76 & 0.00 & 13.034 & 0.056 \\
Si$\:$IV & 1402.77 & 4118.72 & 1.936137 & 0.000000 & 5.27 & 0.00 & 12.995 & 0.214 \\
H$\:$I & 1215.67 & 4118.82 & 2.388111 & 0.000030 & 12.35 & 2.08 & 13.065 & 0.117 \\
H$\:$I & 1215.67 & 4119.53 & 2.388697 & 0.000010 & 19.36 & 1.28 & 13.730 & 0.035 \\
H$\:$I & 1215.67 & 4123.34 & 2.391830 & 0.000048 & 33.36 & 6.13 & 13.134 & 0.068 \\
\hline  
\end{tabular}
\end{minipage}
\end{table*}
\newpage
\setcounter{table}{3}
\begin{table*}
\begin{minipage}{120mm}
\caption{\rm Absorption line parameter list}
\begin{tabular}{lrrrrrrrr}
\hline  
ID & $\lambda_{\rm rest}$ & $\lambda_{\rm obs}$ & $z$ & $\sigma_{z}$ & $b$  & $\sigma_{b}$ & Log$N$  & $\sigma_{{\rm Log}N}$\\
\hline  
H$\:$I & 1215.67 & 4201.59 & 2.456196 & 0.000007 & 20.86 & 0.84 & 13.364 & 0.016 \\
Si$\:$III & 1206.50 & 4203.22 & 2.483818 & 0.000025 & 13.50 & 0.79 & 14.224 & 0.295 \\
Si$\:$III & 1206.50 & 4203.74 & 2.484244 & 0.000041 & 20.32 & 2.22 & 13.278 & 0.097 \\
H$\:$I & 1215.67 & 4207.29 & 2.460888 & 0.000051 & 50.90 & 6.16 & 13.084 & 0.045 \\
H$\:$I & 1215.67 & 4209.72 & 2.462880 & 0.000033 & 23.53 & 4.06 & 12.758 & 0.063 \\
H$\:$I & 1215.67 & 4211.76 & 2.464566 & 0.000006 & 30.67 & 0.78 & 14.123 & 0.021 \\
H$\:$I & 1215.67 & 4214.71 & 2.466985 & 0.000015 & 23.07 & 1.72 & 13.328 & 0.030 \\
\hline  
\end{tabular}
\end{minipage}
\end{table*}
\newpage
\setcounter{table}{3}
\begin{table*}
\begin{minipage}{120mm}
\caption{\rm Absorption line parameter list}
\begin{tabular}{lrrrrrrrr}
\hline  
ID & $\lambda_{\rm rest}$ & $\lambda_{\rm obs}$ & $z$ & $\sigma_{z}$ & $b$  & $\sigma_{b}$ & Log$N$  & $\sigma_{{\rm Log}N}$\\
\hline  
N$\:$I & 1199.55 & 4348.59 & 2.625185 & 0.000013 & 19.12 & 1.51 & 14.704 & 0.072 \\
N$\:$I & 1199.55 & 4348.82 & 2.625376 & 0.000041 & 33.77 & 2.34 & 14.387 & 0.153 \\
N$\:$I & 1199.55 & 4349.12 & 2.625624 & 0.000006 & 7.64 & 0.74 & 14.318 & 0.050 \\
N$\:$I & 1200.22 & 4351.04 & 2.625185 & 0.000013 & 19.12 & 1.51 & 14.704 & 0.072 \\
N$\:$I & 1200.22 & 4351.27 & 2.625376 & 0.000041 & 33.77 & 2.34 & 14.387 & 0.153 \\
N$\:$I & 1200.22 & 4351.56 & 2.625624 & 0.000006 & 7.64 & 0.74 & 14.318 & 0.050 \\
H$\:$I & 1215.67 & 4352.60 & 2.580412 & 0.000021 & 36.93 & 1.71 & 13.831 & 0.032 \\
\hline  
\end{tabular}
\end{minipage}
\end{table*}
\newpage
\setcounter{table}{3}
\begin{table*}
\begin{minipage}{120mm}
\caption{\rm Absorption line parameter list}
\begin{tabular}{lrrrrrrrr}
\hline  
ID & $\lambda_{\rm rest}$ & $\lambda_{\rm obs}$ & $z$ & $\sigma_{z}$ & $b$  & $\sigma_{b}$ & Log$N$  & $\sigma_{{\rm Log}N}$\\
\hline  
Si$\:$II & 1526.71 & 4481.28 & 1.935259 & 0.000034 & 11.16 & 3.95 & 13.086 & 0.161 \\
H$\:$I & 1215.67 & 4483.04 & 2.687711 & 0.000161 & 78.46 & 23.65 & 12.863 & 0.100 \\
H$\:$I & 1215.67 & 4485.30 & 2.689567 & 0.000045 & 21.08 & 3.30 & 13.850 & 0.130 \\
H$\:$I & 1215.67 & 4485.92 & 2.690076 & 0.000025 & 53.35 & 1.26 & 14.325 & 0.030 \\
H$\:$I & 1215.67 & 4488.50 & 2.692196 & 0.000062 & 34.84 & 6.63 & 13.230 & 0.083 \\
H$\:$I & 1215.67 & 4489.50 & 2.693025 & 0.000040 & 26.24 & 3.18 & 14.058 & 0.076 \\
N$\:$V & 1238.82 & 4490.14 & 2.624522 & 0.000007 & 6.64 & 0.90 & 13.246 & 0.039 \\
\hline  
\end{tabular}
\end{minipage}
\end{table*}
\newpage
\setcounter{table}{3}
\begin{table*}
\begin{minipage}{120mm}
\caption{\rm Absorption line parameter list}
\begin{tabular}{lrrrrrrrr}
\hline  
ID & $\lambda_{\rm rest}$ & $\lambda_{\rm obs}$ & $z$ & $\sigma_{z}$ & $b$  & $\sigma_{b}$ & Log$N$  & $\sigma_{{\rm Log}N}$\\
\hline  
H$\:$I & 1215.67 & 4555.48 & 2.747295 & 0.000030 & 8.99 & 5.44 & 12.065 & 0.286 \\
H$\:$I & 1215.67 & 4556.82 & 2.748399 & 0.000009 & 24.56 & 1.02 & 13.374 & 0.020 \\
H$\:$I & 1215.67 & 4558.45 & 2.749737 & 0.000050 & 70.61 & 4.54 & 13.675 & 0.031 \\
H$\:$I & 1215.67 & 4558.90 & 2.750113 & 0.000010 & 12.46 & 1.00 & 13.333 & 0.044 \\
H$\:$I & 1215.67 & 4559.37 & 2.750500 & 0.000056 & 22.11 & 6.56 & 12.911 & 0.171 \\
H$\:$I & 1215.67 & 4560.90 & 2.751753 & 0.000003 & 20.84 & 0.41 & 13.904 & 0.012 \\
H$\:$I & 1215.67 & 4563.09 & 2.753555 & 0.000120 & 100.00 & 19.56 & 13.040 & 0.064 \\
\hline  
\end{tabular}
\end{minipage}
\end{table*}
\newpage
\newpage
\setcounter{table}{3}
\begin{table*}
\begin{minipage}{120mm}
\caption{\rm Absorption line parameter list}
\begin{tabular}{lrrrrrrrr}
\hline  
ID & $\lambda_{\rm rest}$ & $\lambda_{\rm obs}$ & $z$ & $\sigma_{z}$ & $b$  & $\sigma_{b}$ & Log$N$  & $\sigma_{{\rm Log}N}$\\
\hline  
Si$\:$II & 1190.42 & 4655.47 & 2.910793 & 0.000002 & 6.02 & 0.30 & 12.414 & 0.018 \\
Si$\:$II & 1190.42 & 4655.85 & 2.911111 & 0.000001 & 10.40 & 0.19 & 12.853 & 0.006 \\
Si$\:$II & 1190.42 & 4656.21 & 2.911411 & 0.000011 & 3.70 & 1.87 & 11.272 & 0.095 \\
H$\:$I & 1215.67 & 4657.55 & 2.831258 & 0.000031 & 16.23 & 4.15 & 12.539 & 0.119 \\
H$\:$I & 1215.67 & 4658.32 & 2.831891 & 0.000044 & 84.19 & 3.93 & 13.537 & 0.020 \\
H$\:$I & 1215.67 & 4662.54 & 2.835365 & 0.000005 & 63.11 & 1.10 & 15.715 & 0.056 \\
Si$\:$II & 1193.29 & 4665.34 & 2.909643 & 0.000003 & 8.10 & 0.50 & 12.476 & 0.035 \\
\hline  
\end{tabular}
\end{minipage}
\end{table*}
\newpage
\setcounter{table}{3}
\begin{table*}
\begin{minipage}{120mm}
\caption{\rm Absorption line parameter list}
\begin{tabular}{lrrrrrrrr}
\hline  
ID & $\lambda_{\rm rest}$ & $\lambda_{\rm obs}$ & $z$ & $\sigma_{z}$ & $b$  & $\sigma_{b}$ & Log$N$  & $\sigma_{{\rm Log}N}$\\
\hline  
H$\:$I & 1215.67 & 4768.45 & 2.922484 & 0.000010 & 14.87 & 1.15 & 12.841 & 0.028 \\
H$\:$I & 1215.67 & 4774.49 & 2.927459 &  0.000016 &  34.91  &   1.62 & 13.320  &  0.018\\
Ni$\:$II & 1317.22 & 4775.07 & 2.625111 &  0.000296 &   31.35 &   10.65 &  13.310 &    0.527\\
Ni$\:$II & 1317.22 & 4775.30 &2.625286 & 0.000016 &  17.20 &   3.17 & 13.532 &   0.275\\
Ni$\:$II & 1317.22 & 4775.77 &2.625646 & 0.000003  &  5.33  &  0.62 & 13.168 &   0.063\\
Ni$\:$II & 1317.22 & 4776.04 &2.625847 &  0.000046 &  15.20  &   4.54 & 12.921 & 0.159\\
H$\:$I & 1215.67 & 4777.80 & 2.930174 & 0.000005 & 29.71 & 0.52 & 14.054 & 0.010 \\
\hline  
\end{tabular}
\end{minipage}
\end{table*}
\newpage
\setcounter{table}{3}
\begin{table*}
\begin{minipage}{120mm}
\caption{\rm Absorption line parameter list}
\begin{tabular}{lrrrrrrrr}
\hline  
ID & $\lambda_{\rm rest}$ & $\lambda_{\rm obs}$ & $z$ & $\sigma_{z}$ & $b$  & $\sigma_{b}$ & Log$N$  & $\sigma_{{\rm Log}N}$\\
\hline  
H$\:$I & 1215.67 & 4864.71 & 3.001668 & 0.000012 & 21.57 & 2.09 & 13.080 & 0.112 \\
H$\:$I & 1215.67 & 4865.28 & 3.002140 & 0.000271 & 40.55 & 15.24 & 12.708 & 0.276 \\
H$\:$I & 1215.67 & 4868.81 & 3.005042 & 0.000094 & 93.08 & 13.04 & 12.824 & 0.053 \\
H$\:$I & 1215.67 & 4870.75 & 3.006637 & 0.000064 & 32.89 & 3.43 & 14.099 & 0.334 \\
H$\:$I & 1215.67 & 4871.32 & 3.007104 & 0.000568 & 43.09 & 24.65 & 13.782 & 0.704 \\
H$\:$I & 1215.67 & 4873.02 & 3.008506 & 0.000011 & 19.09 & 1.88 & 12.934 & 0.091 \\
H$\:$I & 1215.67 & 4873.32 & 3.008747 & 0.000148 & 67.53 & 10.46 & 13.387 & 0.102 \\
\hline  
\end{tabular}
\end{minipage}
\end{table*}
\newpage
\setcounter{table}{3}
\begin{table*}
\begin{minipage}{120mm}
\caption{\rm Absorption line parameter list}
\begin{tabular}{lrrrrrrrr}
\hline  
ID & $\lambda_{\rm rest}$ & $\lambda_{\rm obs}$ & $z$ & $\sigma_{z}$ & $b$  & $\sigma_{b}$ & Log$N$  & $\sigma_{{\rm Log}N}$\\
\hline  
O$\:$I & 1302.17 & 5091.69 & 2.910164 & 0.000000 & 10.33 & 0.00 & 14.837 & 0.043 \\
O$\:$I & 1302.17 & 5091.86 & 2.910289 & 0.000000 & 15.14 & 0.00 & 14.229 & 0.116 \\
O$\:$I & 1302.17 & 5092.51 & 2.910793 & 0.000000 & 7.97 & 0.00 & 12.460 & 0.187 \\
O$\:$I & 1302.17 & 5092.93 & 2.911111 & 0.000000 & 13.78 & 0.00 & 13.002 & 0.067 \\
Si$\:$II & 1304.37 & 5099.62 & 2.909643 & 0.000003 & 8.10 & 0.50 & 12.476 & 0.035 \\
Si$\:$II & 1304.37 & 5100.29 & 2.910154 & 0.000011 & 31.31 & 1.33 & 13.266 & 0.016 \\
Si$\:$II & 1304.37 & 5100.30 & 2.910164 & 0.000003 & 7.79 & 0.22 & 14.011 & 0.034 \\
\hline  
\end{tabular}
\end{minipage}
\end{table*}
\newpage
\setcounter{table}{3}
\clearpage
\begin{table*}
\begin{minipage}{120mm}
\caption{\rm Absorption line parameter list}
\begin{tabular}{lrrrrrrrr}
\hline  
ID & $\lambda_{\rm rest}$ & $\lambda_{\rm obs}$ & $z$ & $\sigma_{z}$ & $b$  & $\sigma_{b}$ & Log$N$  & $\sigma_{{\rm Log}N}$\\
\hline  
C$\:$IV & 1550.77 & 5406.92 & 2.486600 & 0.000004 & 2.41 & 1.00 & 12.415 & 0.061 \\
Al$\:$III & 1854.72 & 5443.90 & 1.935163 & 0.000003 & 5.93 & 0.45 & 12.219 & 0.020 \\
Si$\:$IV & 1393.76 & 5449.90 & 2.910228 & 0.000007 & 27.87 & 0.85 & 12.990 & 0.011 \\
Si$\:$IV & 1393.76 & 5450.74 & 2.910827 & 0.000003 & 7.62 & 0.38 & 12.731 & 0.021 \\
Si$\:$IV & 1393.76 & 5451.29 & 2.911224 & 0.000006 & 21.18 & 0.71 & 13.073 & 0.011 \\
Si$\:$IV & 1393.76 & 5452.68 & 2.912224 & 0.000023 & 14.36 & 2.82 & 12.051 & 0.065 \\
Si$\:$IV & 1393.76 & 5453.42 & 2.912751 & 0.000011 & 13.01 & 1.38 & 12.306 & 0.035 \\
\hline  
\end{tabular}
\end{minipage}
\end{table*}
\newpage
\setcounter{table}{3}
\begin{table*}
\begin{minipage}{120mm}
\caption{\rm Absorption line parameter list}
\begin{tabular}{lrrrrrrrr}
\hline  
ID & $\lambda_{\rm rest}$ & $\lambda_{\rm obs}$ & $z$ & $\sigma_{z}$ & $b$  & $\sigma_{b}$ & Log$N$  & $\sigma_{{\rm Log}N}$\\
\hline  
C$\:$IV & 1550.77 & 5886.80 & 2.796049 & 0.000046 & 29.07 & 5.47 & 12.554 & 0.067 \\
Na$\:$I & 5891.58 & 5890.67 & -0.000154 & 0.000000 & 4.18 & 0.22 & 11.920 & 0.012 \\
Na$\:$I & 5891.58 & 5891.07 & -0.000087 & 0.000001 & 1.78 & 0.17 & 12.121 & 0.046 \\
Na$\:$I & 5891.58 & 5891.27 & -0.000052 & 0.000001 & 3.16 & 0.34 & 12.044 & 0.015 \\
Na$\:$I & 5897.56 & 5896.65 & -0.000154 & 0.000000 & 4.18 & 0.22 & 11.920 & 0.012 \\
Na$\:$I & 5897.56 & 5897.04 & -0.000087 & 0.000001 & 1.78 & 0.17 & 12.121 & 0.046 \\
Na$\:$I & 5897.56 & 5897.25 & -0.000052 & 0.000001 & 3.16 & 0.34 & 12.044 & 0.015 \\
\hline  
\end{tabular}
\end{minipage}
\end{table*}
\newpage
\setcounter{table}{3}
\begin{table*}
\begin{minipage}{120mm}
\caption{\rm Absorption line parameter list}
\begin{tabular}{lrrrrrrrr}
\hline  
ID & $\lambda_{\rm rest}$ & $\lambda_{\rm obs}$ & $z$ & $\sigma_{z}$ & $b$  & $\sigma_{b}$ & Log$N$  & $\sigma_{{\rm Log}N}$\\
\hline  
C$\:$IV & 1548.20 & 6063.19 & 2.916300 & 0.000007 & 16.83 & 0.81 & 12.916 & 0.017 \\
C$\:$IV & 1550.77 & 6065.48 & 2.911272 & 0.000016 & 49.83 & 1.87 & 13.722 & 0.018 \\
C$\:$IV & 1550.77 & 6065.69 & 2.911406 & 0.000022 & 16.19 & 2.89 & 13.087 & 0.097 \\
C$\:$IV & 1550.77 & 6066.03 & 2.911626 & 0.000003 & 3.81 & 0.77 & 12.777 & 0.070 \\
C$\:$IV & 1550.77 & 6070.93 & 2.914786 & 0.000011 & 10.76 & 1.29 & 12.595 & 0.043 \\
C$\:$IV & 1550.77 & 6071.46 & 2.915124 & 0.000003 & 10.54 & 0.42 & 13.283 & 0.015 \\
C$\:$IV & 1550.77 & 6072.07 & 2.915521 & 0.000014 & 16.99 & 1.71 & 12.832 & 0.035 \\
\hline  
\end{tabular}
\end{minipage}
\end{table*}

\end{document}